\documentclass[final, journal, twocolumn]{IEEEtran}

\usepackage{threeparttable} % to put notes at the end of a table
\usepackage{graphicx}
\usepackage{picinpar}
\usepackage{rotating}

\usepackage{psfrag}
\usepackage{placeins}
\usepackage{times}
\usepackage{amsmath}
\usepackage{amsfonts}
\usepackage{subfigure}
\usepackage{cite}
\usepackage{amsthm}

\newtheorem{proposition}{Proposition}

\newtheorem{conjecture}{Conjecture}

\usepackage{pifont}
\usepackage{amssymb}
\usepackage{setspace}
\usepackage{changebar}
\usepackage{flafter}
\DeclareGraphicsExtensions{.eps,.pdf}
\usepackage{booktabs}
\usepackage{tabularx}
\usepackage{array}

\newcommand{\PreserveBackslash}[1]{\let\temp=\\#1\let\\=\temp}
\newcolumntype{C}[1]{>{\PreserveBackslash\centering}p{#1}}
\newcolumntype{R}[1]{>{\PreserveBackslash\raggedleft}p{#1}}
\newcolumntype{L}[1]{>{\PreserveBackslash\raggedright}p{#1}}

\usepackage{mdwmath}
\usepackage{mdwtab}
\usepackage{dblfloatfix}
\usepackage{morefloats}
\usepackage{url}

\usepackage{lineno}
\setpagewiselinenumbers
\modulolinenumbers[1]

\begin{document}
\bibliographystyle{../../../../shaoshi_bib/IEEEtran}
\title{Approximate Bayesian Probabilistic Data Association Aided Iterative Detection for MIMO Systems Using Arbitrary $M$-ary Modulation}
\author{Shaoshi~Yang, Li~Wang, Tiejun~Lv,~\IEEEmembership{Senior~Member,~IEEE}, and~Lajos~Hanzo,~\IEEEmembership{Fellow,~IEEE}
% Shaoshi~Yang,~\IEEEmembership{Student Member,~IEEE}, Li~Wang,~\IEEEmembership{Member,~IEEE}, and~Lajos~Hanzo,~\IEEEmembership{Fellow,~IEEE}
\thanks{
% Copyright (c) 2011 IEEE. Personal use of this material is permitted. However, permission to use this material for any other purposes must
% be obtained from the IEEE by sending a request to pubs-permissions@ieee.org.
Copyright (c) 2012 IEEE. Personal use of this material is permitted. However, permission to use this material for any other purposes must be obtained from the IEEE by sending a request to pubs-permissions@ieee.org. 

The financial support of the China Scholarship Council (CSC), of the
Research Councils UK (RCUK) under the India-UK Advanced Technology
Center (IU-ATC), and of the EU under the auspices of the Concerto
project is gratefully acknowledged.

S. Yang and L. Hanzo are with the School of Electronics and Computer Science,
University of Southampton, Southampton, SO17 1BJ, UK (e-mail: \{sy7g09, lh\}@ecs.soton.ac.uk).

L. Wang was with the School of Electronics and Computer
Science, University of Southampton, Southampton, SO17 1BJ, UK, and now he is with the R\&D center of Huawei Technologies in Stockholm, Sweden (e-mail: leo.li.wang@huawei.com). 

T. Lv is with the School of Information and Communication
Engineering, Beijing University of Posts and Telecommunications,
Beijing, 100876, China (e-mail: lvtiejun@bupt.edu.cn).
}
}

\markboth{This paper has been published in IEEE Transactions on Vehicular Technology, vol. 62, no. 3, March, 2013.}%
{Shell \MakeLowercase{\textit{et al.}}: Bare Demo of IEEEtran.cls
for Journals}

\maketitle

% \linenumbers
% \vspace{-1.5cm}
\begin{abstract}
% Capacity-approaching forward error correction (FEC) codes and
% multiple-antenna techniques have become essential components of
% advanced wireless communication systems capable of achieving an infinitesimally low error rate, despite their high throughput. However, the optimal
% joint detector/decoder for this system is computationally
% prohibitive.
In this paper, the issue of designing an iterative detection and decoding (IDD) aided
receiver relying on the low-complexity probabilistic data association
(PDA) method,
% \begin{changebar}which typically operates purely in the probability-domain,\end{changebar}
is addressed for turbo-coded
multiple-input--multiple-output (MIMO) systems using general $M$-ary modulations. We demonstrate
that the classic candidate-search aided bit-based
extrinsic log-likelihood ratio (LLR) calculation method is not applicable to
the family of PDA-based detectors. Additionally, we reveal that in contrast to the interpretation in the existing literature,
the output symbol probabilities of existing PDA algorithms are not the true \textit{a posteriori} probabilities (APPs), but rather the normalized
symbol likelihoods. Therefore, the classic relationship,
where the extrinsic LLRs are given by subtracting the \textit{a
priori} LLRs from the \textit{a posteriori} LLRs does not hold for the
existing PDA-based detectors.
Motivated by these revelations, we conceive a new approximate Bayesian theorem based
logarithmic-domain PDA (AB-Log-PDA) method,
% which is capable of exploiting the \textit{a priori} soft information feedback of the FEC decoder. Furthermore, we propose
and unveil the technique of calculating bit-based extrinsic LLRs for the AB-Log-PDA, which facilitates the employment of the AB-Log-PDA in a simplified IDD receiver structure. Additionally, we demonstrate that we may dispense with inner iterations within the AB-Log-PDA in the context of IDD receivers.
% \begin{changebar}where the extrinsic LLRs of the AB-Log-PDA
% can be approximated by its pro forma \textit{a posteriori} LLRs. \end{changebar}
Our complexity analysis and numerical results recorded for Nakagami-$m$ fading channels demonstrate that the proposed AB-Log-PDA based
IDD scheme is capable of achieving
a comparable performance to that of the
optimal MAP detector
based IDD receiver,
% regardless of the number of MIMO elements,
while imposing a significantly lower computational complexity in the scenarios considered.
% This is because the proposed AB-Log-PDA method integrated
% into the IDD receiver is capable of exploiting the soft information of the FEC decoder and hence achieves an improved accuracy
% in modelling the inter-antenna interference (IAI).
%  The effectiveness of the proposed IDD schemes are demonstrated by both
% the numerical simulations and the semi-analytical extrinsic information transfer (EXIT)
% chart analysis.
\end{abstract}


\makeatletter
\def\hlinewd#1{%
  \noalign{\ifnum0=`}\fi\hrule \@height #1 \futurelet
   \reserved@a\@xhline}
\makeatother

\IEEEpeerreviewmaketitle

\section{Introduction}
\IEEEPARstart{W}{hen}
% The past two decades have witnessed a series of
% breakthroughs in the wireless research community, such as multiple-input--multiple-output (MIMO) techniques \cite{Foschini:MIMO, Tarokh:Space-Time_code},
% and near-capacity forward error
% correction (FEC) codes, as exemplified by turbo codes \cite{Berrou:Turbo_coding_conference}
% % \cite{Berrou:Turbo_coding_conference, Berrou:Turbo_coding_journal}
% and low-density parity-check (LDPC) codes \cite{Gallager:LDPC_code}.
% The former significantly improves the capacity of a wireless communication system, while the latter represents a great step towards near error-free
% wireless transmission.
conceiving
advanced wireless systems using both multiple-input--multiple-output (MIMO) techniques \cite{Foschini:MIMO, Tarokh:Space-Time_code} and near-capacity
forward error correction (FEC) codes \cite{Berrou:Turbo_coding_conference,Gallager:LDPC_code,Breiling2000:super_trellis_turbo}
to simultaneously achieve a high throughput and an infinitesimally low
error rate, one of the major challenges is the potentially prohibitive computational complexity at receiver.
The iterative detection and decoding (IDD) \cite{Hagenauer1997:turbo_principle}, which was inspired by decoding of concatenated
codes \cite{Berrou:Turbo_coding_conference, Hagenauer:iterative_decoding_concatenated_codes},
is capable of achieving a
near-optimum performance at a significantly lower complexity than the maximum-likelihood sequence estimator based optimal joint
 detector/decoder \cite{Hochwald:SD_near_capacity,Giallorenzi:optimum_joint_receiver}.
% closely
% approximating the performance of the optimal joint
% detector/decoder.
% When exploiting the ``turbo principle'' \cite{Hagenauer1997:turbo_principle}, the individual decoding modules
% must be capable of both accepting and generating probabilities or soft
% values, where the so-called extrinsic part of the soft output of one
% decoder is delivered to the other decoder as its \textit{a priori}
% information.
%
% As a benefit, a good IDD design is capable of achieving a
% near-optimum performance at a significantly lower complexity than the
% optimal joint detector/decoder \cite{Hochwald:SD_near_capacity}.
%
Even so, the computational complexity imposed by the IDD might
remain the limiting factor in practical applications.

% The existing
% reduced-complexity IDD receivers are typically relying on the list
% sphere decoding (LSD) algorithm \cite{Hochwald:SD_near_capacity} and
% the minimum mean-square error criterion based interference
% cancellation (MMSE-IC) algorithm
% \cite{Wang_Xiaodong_1999:iterative_detection}. The LSD based IDD
% scheme reduces the complexity compared to the optimal maximum
% \textit{a posteriori} (MAP) based scheme, but it sacrifices the
% performance and is not efficient for low signal-to-noise ratio (SNR)
% scenarios. The MMSE-IC aided scheme imposes lower complexity, but
% its achievable performance is much worse than that of the MAP and
% the LSD based schemes.
% In particular, an iterative receiver using the MMSE filtering based soft interference cancellation was
% proposed by X. Wang and H. V. Poor in \cite{Wang_Xiaodong_1999:iterative_detection} for multiuser
% detection in a convolutionally coded CDMA system, and more recently,
% B. M. Hochwald and S. ten Brink elaborated on how to achieve
% near-capacity communication on a multiple-antenna channel relying on
% the list sphere decoder (LSD) aided iterative receiver \cite{Hochwald:SD_near_capacity}.

The probabilistic data association (PDA) filtering method was
highly successful for target tracking problems
\cite{Shalom:PDA_original,Shalom2009:PDA_filter} in radar systems. In digital communications, it may also be developed
to a soft-input--soft-output (SISO)
% reduced-complexity design alternative of MAP decoders/detectors/equalizers
% \cite{Luo:PDA_Sync_CDMA, Pham:GPDA, Liu:CPDA-apx, Jia:CPDA,
% Shaoshi2011:B_PDA, Shaoshi2011:DPDA}.
% The PDA method has several
% attractive properties \cite{Shaoshi2011:B_PDA}. Firstly, it may
% achieve a near-optimal detection performance in high-throughput MIMO
% or CDMA systems having a large number of parallel streams, provided
% that the channel is not rank-deficient. Secondly, it has a low
% worst-case complexity which increases no faster than
% $\mathcal{O}\left( {{\mathcal{N}}^3 } \right)$ for all
% signal-to-noise ratio (SNR) values, where $\mathcal{N}$ is the
% number of parallel streams. Thirdly, it is inherently a
% soft-input--soft-output (SISO) algorithm, which is eminently
% applicable for combination with FEC coding.
% As an
efficient interference-modelling method \cite{Luo:PDA_Sync_CDMA, Pham:GPDA, Liu:CPDA-apx, Jia:CPDA,
Shaoshi2011:B_PDA, Shaoshi2011:DPDA}. The key feature of PDA is
the repeated conversion of a multimodal Gaussian mixture probability
to a single multivariate Gaussian distribution. Therefore, the accuracy of the
Gaussian approximation dominates the attainable performance.
In uncoded MIMO systems using quadrature amplitude modulation (QAM), the quality of the Gaussian approximation in PDA may be
improved by transforming the symbol-based model into a bit-based model, which in effect increases the length of the effective
transmitted signal vector by the number of bits per symbol, and reduces the effective constellation
to a binary constellation \cite{Shaoshi2011:B_PDA}. With regard to improving the quality of the Gaussian approximation in FEC-coded MIMO systems, we
benefit from having an increased exploitable degree
of freedom. For example, the soft information gleaned from the output of the FEC decoder tends to be more
reliable than the output symbol probabilities of the PDA detector itself. Therefore the FEC decoder's soft output would facilitate
a more accurate modelling of the inter-antenna interference (IAI). The iterative receiver proposed in \cite{Cai2006:iterative_PDA} was
essentially a maximum \textit{a posteriori} (MAP) detection aided IDD scheme, employing the hard-output PDA detector for generating the candidate-search list,
hence it did not solve the problem of interest to us.

There are several particular challenges that render the IDD
design using PDA less straightforward than it seems to be.
\textit{Firstly}, to the best of our knowledge, all the existing PDA detectors conceived for uncoded systems operate purely in the
probability-domain\cite{Luo:PDA_Sync_CDMA, Pham:GPDA, Liu:CPDA-apx, Jia:CPDA,
Shaoshi2011:B_PDA, Shaoshi2011:DPDA}, which results in a poor numerical stability in IDD scenarios, hence potentially leading to a degraded
performance.
\textit{Secondly}, it is unclear how to produce the correct bit-based extrinsic log-likelihood ratios (LLRs) required by the
concatenated outer FEC decoder. Conventionally, the output symbol probabilities of the existing PDA algorithms were interpreted as
the \textit{a posteriori} probabilities (APPs)\cite{Luo:PDA_Sync_CDMA, Pham:GPDA, Liu:CPDA-apx, Jia:CPDA,
Shaoshi2011:B_PDA, Shaoshi2011:DPDA}. Hence, one may assume that
a natural way of generating the bit-based extrinsic LLRs is to subtract the bit-based \textit{a priori} LLRs
from the bit-based \textit{a posteriori} LLRs generated from the output probabilities of the PDA algorithms. However, we
will show that this classic relationship no longer holds
if we still treat the output symbol probabilities of the
existing PDA algorithms as APPs in the context of IDD receivers.
\textit{Thirdly}, the existing PDA algorithms\cite{Luo:PDA_Sync_CDMA, Pham:GPDA, Liu:CPDA-apx, Jia:CPDA,
Shaoshi2011:B_PDA, Shaoshi2011:DPDA} have
an inherently self-iterative structure, where the estimated symbol
probabilities are delivered to the next
\textit{inner iteration} after the current inner iteration is
completed. Then, the question of how to deal with the inner iterations of the existing PDAs in the context of IDD receivers arises.

Against this backdrop,
% in this paper we aim for designing a
% low-complexity IDD scheme relying on the PDA approach for FEC-coded
% MIMO systems using general $M$-ary modulations.
the main contributions of this paper are as follows.

1) We present an analysis of the interference-plus-noise
distribution for the MIMO signal model, which sheds light on
the fundamental principles of the PDA algorithms from a new perspective.

2) We propose an approximate Bayesian theorem based
logarithmic-domain PDA (AB-Log-PDA) MIMO detector for IDD aided MIMO systems employing arbitrary $M$-ary modulations, which has not been reported before. The proposed AB-Log-PDA enjoys better numerical stability and accuracy, hence it is better suited for
iterative detection than the existing probability-domain PDA detectors conceived for uncoded systems \cite{Luo:PDA_Sync_CDMA, Pham:GPDA, Liu:CPDA-apx, Jia:CPDA, Shaoshi2011:B_PDA, Shaoshi2011:DPDA}.

% 3) Furthermore, we show that the estimated symbol APPs output by the
% AB-Log-PDA, and also by the existing PDAs, are actually not the ``true''
% symbol APPs, but some sort of ``nominal'' symbol APPs. As
% a result, given these symbol APPs, the bit-based extrinsic LLRs
% cannot be produced according to the classic relationship, where the
% extrinsic LLRs may be obtained by subtracting the \textit{a priori}
% LLRs from the \textit{a posteriori} LLRs, nor by the
% vector-conditioned likelihood function based candidate-search
% approach. Nonetheless, the symbol
% probabilities output by the PDA based methods have been referred to as APPs
% \cite{Luo:PDA_Sync_CDMA, Pham:GPDA, Liu:CPDA-apx, Jia:CPDA,
% Shaoshi2011:B_PDA, Shaoshi2011:DPDA}, and the distinctions between these ``nominal APPs'' and
% the ``true APPs'' has never been reported before.

3) In contrast to the conventional interpretations of the mathematical properties of the estimated output symbol probabilities of the PDA algorithms, we will demonstrate that these probabilities do \textit{not} constitute the true symbol APPs, they rather constitute the normalized symbol likelihoods. Owing to this  misinterpretation\footnote{We note however that these output symbol probabilities were treated as APPs without causing any problems in the uncoded systems considered in \cite{Luo:PDA_Sync_CDMA, Pham:GPDA, Liu:CPDA-apx, Jia:CPDA, Shaoshi2011:B_PDA, Shaoshi2011:DPDA}. The differences between these symbol probabilities and the true APPs have never been reported before, because the calculation of the extrinsic LLRs is not required in the context of uncoded
systems.}, it is flawed to produce the bit-based extrinsic LLRs from the output symbol probabilities of the PDA algorithms by using the classic relationship,  where the extrinsic LLRs are given by subtracting the \textit{a priori} LLRs from the \textit{a posteriori} LLRs. We demonstrate furthermore that the classic candidate-search aided bit-based extrinsic LLRs calculation method, which is used for example by the MAP detector and the list sphere decoder, is not applicable to any PDA-based detector. In order to circumvent the above-mentioned problems, we conceive a new technique of producing the bit-based extrinsic LLRs for the proposed AB-Log-PDA, which results in a simplified IDD structure, where the extrinsic LLRs of the AB-Log-PDA are generated by directly transforming the output symbol probabilities into bit-based LLRs, without subtracting the \textit{a priori} LLRs.

4) We reveal that introducing inner iterations into the AB-Log-PDA
actually degrades the achievable performance of the IDD receiver, which is in contrast to the impact of the inner iterations within the FEC-decoder of other types of iterative receivers. The reasons as to why the inner PDA iterations fail to provide BER improvement are investigated and discussed in detail. Notably, we show that the proposed AB-Log-PDA based IDD scheme invoking
no inner iterations within the AB-Log-PDA strikes an attractive
performance versus complexity tradeoff, which compares favorably to
that of the optimal MAP based IDD scheme in both \textit{perfect and imperfect} channel-estimation scenarios, when communicating over
Nakagami-$m$ fading channels. For example, in some scenarios the
performance of the proposed AB-Log-PDA based IDD scheme approaches
that of the MAP-based IDD scheme within 0.5 dB, while imposing a
significantly lower computational complexity.

The remainder of the paper is organized as follows. In Section
\ref{Sec:system_model}, our FEC-coded MIMO system model is introduced, while
in Section \ref{Sec:interference_plus_noise_distribution_analysis},
% the distribution of the interference-plus-noise term of the received
% signal model is analyzed, which sheds light on the fundamental
% principles of the PDA method.
we present our analysis of the interference-plus-noise
distribution for our MIMO signal model, which sheds light on the fundamental principles of the PDA from a new perspective.
In Section
\ref{Sec:PDA_in_uncoded_system}, the proposed AB-Log-PDA is presented, which relies on the \textit{a priori} soft information feedback
gleaned from the FEC decoder.
Then, in
Section \ref{Sec:PDA_in_coded_system} the extrinsic LLR calculation of the AB-Log-PDA is detailed, while our
simulation results and discussions are presented in
Section \ref{Sec:Simulations}. Finally, the paper is concluded in
Section \ref{Sec:conclusions}.

\section{System Model}
\label{Sec:system_model}
\begin{figure}[tbp]
\centering
\includegraphics[width=3.6in]{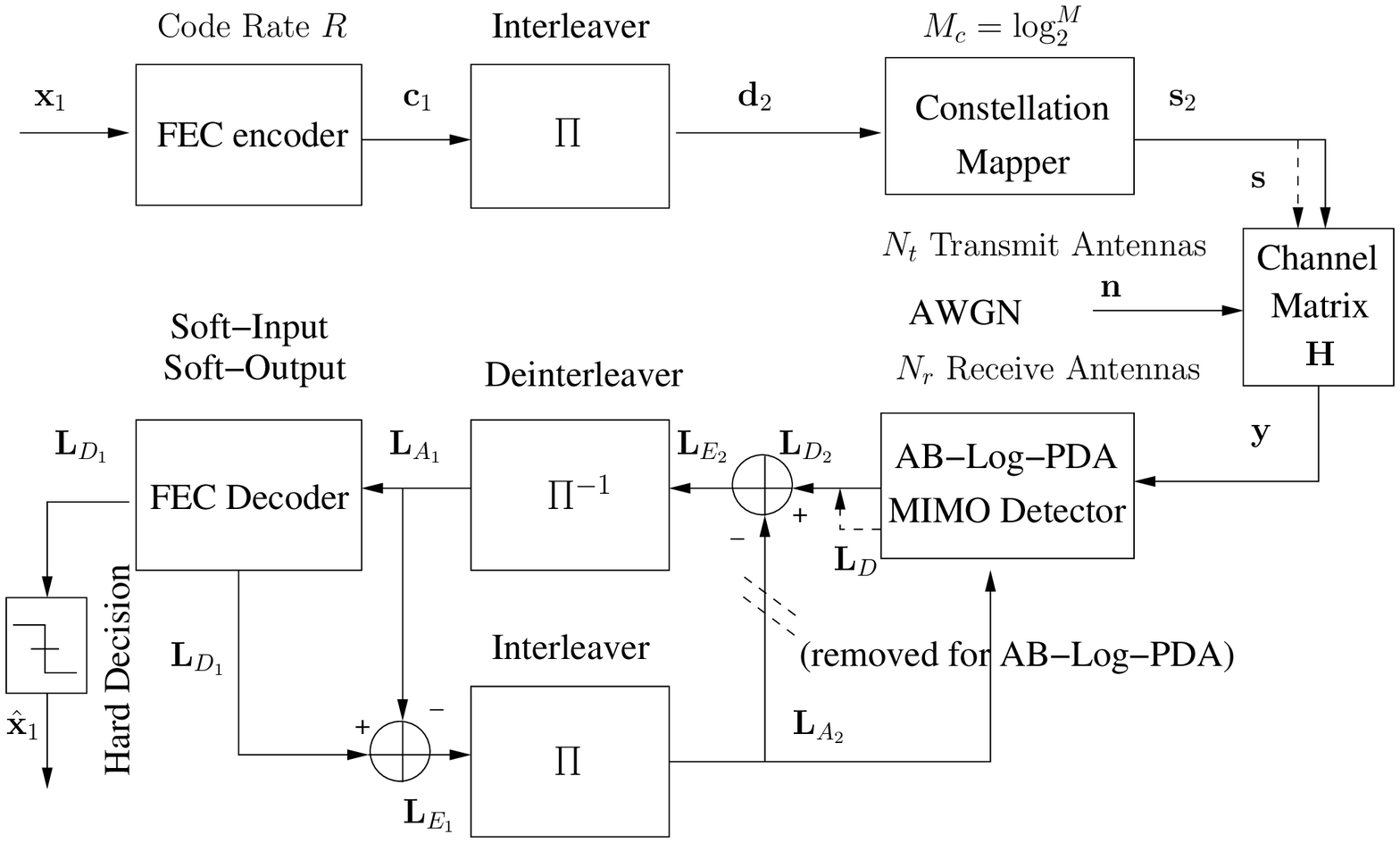}
\caption{The FEC-coded MIMO system with a simplified structure for the AB-Log-PDA based IDD receiver,
where we have ${\bf{L}}_{E_2} = {\bf{L}}_{D_2}$ rather than the
classical ${\bf{L}}_{E_2} = {\bf{L}}_{D_2} - {\bf{L}}_{A_2}$. The subscript ``1'' denotes
the processing modules associated with the outer FEC encoder/decoder, and the subscript ``2'' denotes the processing modules
that are connected with the inner space-time mapper/detector. The arrow with dashed line indicates that $\bf{s}$
and ${\bf{L}}_D$ are the subvectors of ${\bf{s}}_2$ and ${\bf{L}}_{D_2}$, respectively.
% The closed loop of ${\bf{L}}_{E_2}\rightarrow$
% ${\bf{L}}_{A_1}\rightarrow$ ${\bf{L}}_{E_1}\rightarrow$ ${\bf{L}}_{A_2}$ $\rightarrow$ ${\bf{L}}_{E_2}$ represents an global iteration compared
% to the local iteration inside the soft FEC decoder or the soft MIMO detector.
}
\label{fig:SI_PDA_schematic}
\end{figure}
We consider the FEC-coded MIMO system of Fig.~\ref{fig:SI_PDA_schematic}.
At the transmitter, the $(L_f \times 1)$ source-bit
frame ${\bf{x}}_1$ is firstly encoded by a rate $R < 1$ FEC encoder (typically a convolutional code, a turbo code or an LDPC code) into
the $(\frac{L_f}{R} \times 1)$  coded-bit frame ${\bf{c}}_1$. In order to guard against bursty fading,
${\bf{c}}_1$ is then passed through a bit-interleaver. Then the $(\frac{L_f}{R} \times 1)$ interleaver output bit-frame ${\bf{d}}_2$ is mapped
to the $(\frac{L_f}{RM_b} \times 1)$ symbol-frame ${\bf{s}}_2$, with each symbol taken from the $M$-ary modulation
constellation $\mathcal {A}=\{a_1,a_2,\cdots,a_M
\}$, where $M_b = \log_2^M$ is the number of bits per constellation symbol.
Finally, ${\bf{s}}_2$ is transmitted in form of the $(N_t \times 1)$
\textit{vector of symbols} $\bf{s}$ by $N_t\ge 1$ transmit antennas per channel use.
% \footnote{This indicates
% that a single FEC-coded bit-frame
% ${\bf{c}}_1$ is transmitted within $\frac{L_f}{RM_bN_t}$ channel uses.}.
%
% The transmitter of Fig. \ref{fig:SI_PDA_schematic}
% may be regarded as a serial concatenation of the ``outer'' FEC encoder
% and the ``inner'' space-time bit-to-symbol mapper. As also stated in \cite{Hochwald:SD_near_capacity}, the overall iterative
% detection and decoding framework of
% Fig.~\ref{fig:SI_PDA_schematic} has been widely used, and the FEC encoder/decoder is also relatively mature. Therefore,
% the soft MIMO detector becomes the key module that has to be carefully designed to assist the entire system to achieve a high performance
% at a low complexity.

At the output of the fading channel $\bf{H}$, the received $(N_r \times 1)$-element complex-valued baseband signal vector per channel use
is represented by
\begin{equation} \label{eq:MIMO_model}
{\bf{y}=\bf{Hs}+\bf{n}},
\end{equation}
where ${\bf{s}} = [s_1, s_2, \cdots, s_{N_t}]^T$ is normalized by
the component-wise energy constraint $\mathbb{E}({\left | s_i
\right |}^2) = E_{\bf{s}}/N_t$ in order to maintain a total transmit
power $E_{\bf{s}}$ per channel use; and $\bf{n}$ is the $(N_r \times
1)$-element zero-mean complex-valued Gaussian noise vector with a
covariance matrix of $2\sigma^2{\bf{I}}_{N_r}$ where
${\bf{I}}_{N_r}$ represents an $(N_r \times N_r)$-element identity
matrix; and $\bf{H}$ is an $(N_r \times N_t)$ complex-valued matrix
with entries $h_{ji}$  perfectly known to the receiver, $j = 1,
\cdots, N_r$, $i = 1, \cdots, N_t$. In this paper, we assume that
$h_{ji} = r\exp({j\theta})$ is independent and identically
distributed (i.i.d), the phase $\theta$ is uniformly distributed and
independent of the envelope $r$, while $r$ obeys the Nakagami-$m$
distribution having the probability density function (PDF) of
\cite{Proakis:Digtal_Comm_book}
\begin{equation}
\label{Eq:nakagami-m}
p(r) = \frac{2}{\Gamma(m)}\left(\frac{m}{\Omega}\right)^m r^{2m-1}\exp({-mr^2/\Omega}),\; r\ge 0,
\end{equation}
where $\Gamma( \cdot)$ represents the Gamma function, $\Omega
\triangleq \mathbb{E}(r^2)$, and the Nakagami \textit{fading
parameter} is $m \triangleq
{\Omega^2}/{\mathbb{E}[(r^2-\Omega)^2]}$, $m \ge 0.5$. Note that the
Nakagami-$m$ fading model captures a wide range of realistic fading
environments, encompassing the most frequently used Rayleigh and
Rician fading models as special cases. More specifically, the
parameter $m$ indicates the severity of the fading. As $m$ becomes
smaller, the fading effects become more severe. For example, when
$m$ decreases to $0.5$, Eq. (\ref{Eq:nakagami-m}) approaches the
one-sided Gaussian distribution; when $m =1$, Eq.
(\ref{Eq:nakagami-m}) reduces to a Rayleigh PDF, and as
$m\rightarrow\infty$, Eq. (\ref{Eq:nakagami-m}) reduces to a
$\delta$-distribution located at $r = 1$, which corresponds to
imposing no fading on the amplitude of the transmitted signal, but
only a ``pure random phase'' obeying a uniform distribution on the
circle of radius $\sqrt{\Omega}$. In addition to its generalized
nature, the Nakagami-$m$ fading model was shown to fit the
experimental propagation data better than the Rayleigh, Rician and
Lognormal distributions \cite{Aulin1981:Nakagami_channel}.

\section{Interference-Plus-Noise Distribution Analysis}
\label{Sec:interference_plus_noise_distribution_analysis} 
In order
to provide more insight on the fundamental principle underlying the
PDA method, an interference-plus-noise distribution analysis is carried
out in this section.

The received signal model of (\ref{eq:MIMO_model}) may be rewritten
as
\begin{equation} \label{eq:MIMO_Decoupled_Model}
{\bf{y}} = s_i {\bf{h}}_i  + \underbrace{\sum\limits_{k \ne i} { s_k
{\bf{h}}_k}}_{{\bf{u}}_i} + {\bf{n}}  \buildrel \Delta \over = s_i
{\bf{h}}_i  +  \underbrace{{\bf{u}}_i + \bf{n}}_{{\bf{v}}_i},
\end{equation}
where ${\bf{h}}_i$ denotes the $i$-th column of $\bf{H}$, and $s_i$
is the $i$-th symbol of $\bf{s}$, while ${\bf{u}}_i$ is the sum of
$(N_t -1) $ IAI components
contaminating the symbol $s_i$, $i$, $k=1,2,\cdots,N_t$, and
${\bf{v}}_i$ is the interference-plus-noise term for $s_i$.

\begin{figure}[tbp]
\centering
\includegraphics[width=3.5in]{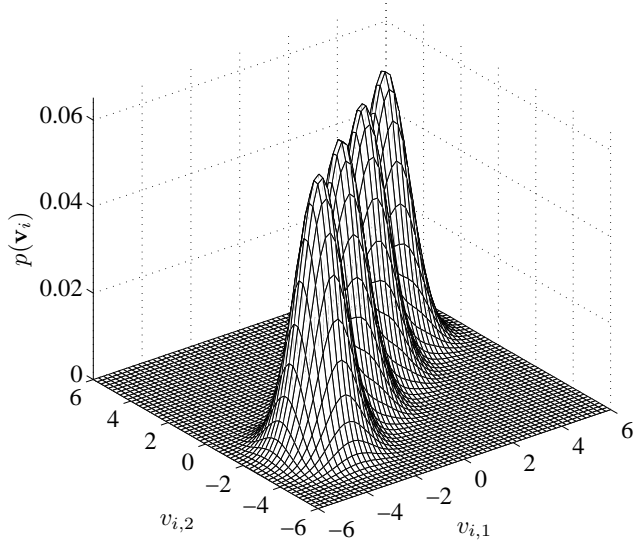}
\caption{The multimodal Gaussian mixture PDF of ${\bf{v}}_i =
[v_{i,1}, v_{i,2}]^T$ for a $N_t = N_r = 2$ MIMO system. For
visualization purpose, the real-valued $4$PAM modulation with
constellation $\mathcal{A} =\{ {-3, -1, 1, 3}\}$ is used. $s_1$ is
assumed to be detected, $s_2$ is the interference signal to $s_1$,
and the real-valued Gaussian channel vectors ${\bf{h}}_1 = [0.8884,
-1.1471]^T$, ${\bf{h}}_2 = [-1.0689, -0.8095]^T$. The possible
interference patterns are $\underline{\bf{u}}_{1,1} = [3.2066,
2.4285]^T$, $\underline{\bf{u}}_{1,2} = [1.0689,   0.8095]^T$,
$\underline{\bf{u}}_{1,3} = [-1.0689,  -0.8095]^T$,
$\underline{\bf{u}}_{1,4} = [-3.2066,   -2.4285]^T$, and a given a
\textit{priori} probability vector ${\bf{P}}_n =
[0.25,0.25,0.25,0.25]$ is used for the possible interference
patterns. The means of the four component Gaussian distributions are
$\underline{\bf{u}}_{1,1}$, $\underline{\bf{u}}_{1,2}$,
$\underline{\bf{u}}_{1,3}$, $\underline{\bf{u}}_{1,4}$,
respectively, and the covariance matrices of the four component
Gaussian distributions are all $\sigma^2\bf{I}$ where $\sigma^2 =
0.631$. } \label{fig:GMM_PDF}
\end{figure}

\begin{figure}[tbp]
\centering
\includegraphics[width=4.0in]{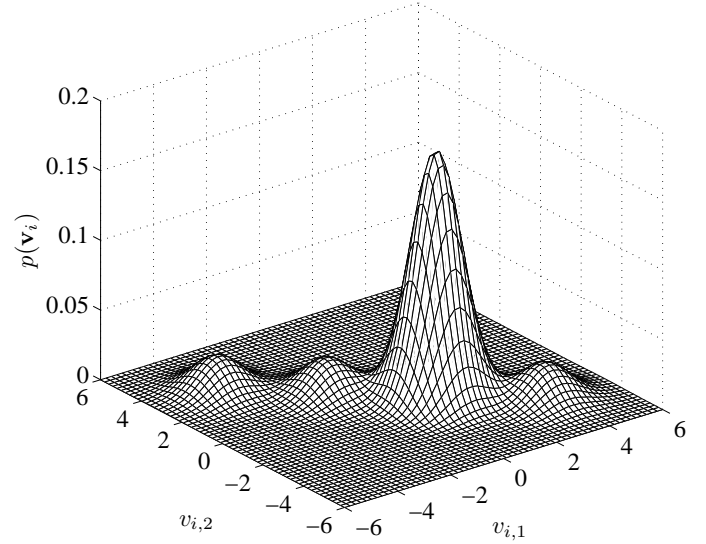}
\caption{The multimodal Gaussian mixture PDF of ${\bf{v}}_i =
[v_{i,1}, v_{i,2}]^T$ for a $N_t = N_r = 2$ MIMO system after
iteration. For visualization purpose, the real-valued $4$PAM
modulation with constellation $\mathcal{A} =\{ {-3, -1, 1, 3}\}$ is
used. $s_1$ is assumed to be detected, $s_2$ is the interference
signal to $s_1$, and the real-valued Gaussian channel vectors
${\bf{h}}_1 = [-3.0292, -0.4570]^T$, ${\bf{h}}_2 = [1.2424,
-1.0667]^T$. The possible interference patterns are
$\underline{\bf{u}}_{1,1} = [-3.7273, 3.2001]^T$,
$\underline{\bf{u}}_{1,2} = [-1.2424,    1.0667]^T$,
$\underline{\bf{u}}_{1,3} = [1.2424,   -1.0667]^T$,
$\underline{\bf{u}}_{1,4} = [3.7273,   -3.2001]^T$, and a given a
\textit{priori} probability vector ${\bf{P}}_n = [0.1,0.1,0.7,0.1]$
is used for the possible interference patterns. The means of the
four component Gaussian distributions are
$\underline{\bf{u}}_{1,1}$, $\underline{\bf{u}}_{1,2}$,
$\underline{\bf{u}}_{1,3}$, $\underline{\bf{u}}_{1,4}$,
respectively, and the covariance matrices of the four component
Gaussian distributions are all $\sigma^2\bf{I}$ where $\sigma^2 =
0.631$. } \label{fig:3D_GMM_after_iteration}
\end{figure}

Note that if ${\bf{u}}_i$ vanishes, Eq.
(\ref{eq:MIMO_Decoupled_Model}) reduces to the classic
single-input--multiple-output interference-free broadcast
channel having a receive diversity order of $N_r$.
Similarly, if we know exactly the distribution
of ${\bf{u}}_i$ or $s_k$, this potentially allows us to mitigate the adverse effects of the IAI.
% \footnote{As mentioned before, ${\bf{h}}_k$ is assumed
% to be perfectly known to the receiver, hence it can be regarded
% deterministic per channel use.}
However, unfortunately, the distribution of ${\bf{u}}_i$ is generally
unknown. A notable exception is, when the number of independent
IAI components is sufficiently high, ${\bf{u}}_i$ approaches a
multivariate Gaussian distribution according to the central limit
theorem,

On the other hand, we observe that the interference term
${\bf{u}}_i$ has a total of $M^{N_t-1}$ possible interference
patterns. Then, the $n$-th legitimate interference pattern imposed by a
given sample of
\begin{equation}
\underline{\bf{s}}_n = [\underline{s}_{1,n},\cdots,
\underline{s}_{k,n}, \cdots, \underline{s}_{N_t,n}]_{k\neq i}^T
\end{equation}
is defined as
\begin{equation}
\underline{\bf{u}}_{i,n} \triangleq \sum\limits_{k\ne i}
\underline{s}_{k,n}{\bf{h}}_k,
\end{equation}
while the corresponding interference-plus-noise pattern is defined as
\begin{equation}
 \underline{\bf{v}}_{i,n} \triangleq
\underline{\bf{u}}_{i,n} + \bf{n},
\end{equation}
where $\underline{s}_{k,n} = a_m\in \mathcal{A}$, $n = 1, 2, \cdots,
M^{N_t - 1}$.  We observe that $\underline{\bf{v}}_{i,n}$ obeys a
multivariate Gaussian distribution with a mean of $\sum\limits_{k
\neq i }\underline{s}_{k,n}{\bf{h}}_k$ and a covariance of
$2\sigma^2{\bf{I}}_{N_r}$ for complex-valued noise, hence the PDF of
$\underline{\bf{v}}_{i,n}$ is formulated as
\begin{equation}
 f(\underline{\bf{v}}_{i,n})  = c\exp \left(-\frac{\lVert\underline{\bf{v}}_{i,n} -
\sum\limits_{k \neq
i}\underline{s}_{k,n}{\bf{h}}_k\rVert^2}{2\sigma^2}\right),
\end{equation}
where
% \begin{eqnarray}
%  c & = & \frac{1}{\pi^{N_r}\det(2\sigma^2{\bf{I}})} \nonumber \\
%    & = & \frac{1}{(2\pi\sigma^2)^{N_r}}.
% \end{eqnarray}
\begin{equation}
 c  =  \frac{1}{\pi^{N_r}\det(2\sigma^2{\bf{I}})}
    =  \frac{1}{(2\pi\sigma^2)^{N_r}}.
\end{equation}
If we assume that the probability of encountering the $n$-th
interference pattern caused by $\underline{{\bf{s}}}_n$ is $P_n$,
then upon jointly considering the distribution of
$\underline{\bf{s}}_n$ and that of $\underline{\bf{v}}_{i,n}$, the
distribution of ${\bf{v}}_i$ may be characterized by the multimodal
Gaussian mixture distribution of
\begin{equation}
\label{eq:true_distribution_of_interference_plus_noise}
 p({\bf{v}}_i) = \sum\limits_{n = 1}^{M^{N_t -1}} {P_n}f(\underline{\bf{v}}_{i,n}),
\end{equation}
where we have $\sum\limits_{n=1}^{M^{N_t -1}} P_n = 1$. Eq.
(\ref{eq:true_distribution_of_interference_plus_noise}) indicates
that the true distribution of ${\bf{v}}_i$ is a weighted average of
a set of component multivariate Gaussian distributions. Note that
Eq.~(\ref{eq:true_distribution_of_interference_plus_noise}) is
\textit{not} the PDF used by the classic optimal MAP detection.
Since $P_n$ is not known beforehand, and the complexity of computing
$p({\bf{v}}_i)$ increases at an exponential rate of
$\mathcal{O}(M^{N_t-1})$, it is infeasible to carry out symbol
detection directly relying on
Eq.~(\ref{eq:true_distribution_of_interference_plus_noise}) for
large-dimensional MIMO systems. Instead, we can resort to the PDA
method to simplify the MIMO detection. An example of a four-modal
Gaussian mixture distribution is shown in Fig. \ref{fig:GMM_PDF} and
Fig. \ref{fig:3D_GMM_after_iteration} in order to illustrate the
fundamental principle of the PDA. More specifically, Fig.
\ref{fig:GMM_PDF} represents the initial distribution of the
interference-plus-noise term ${\bf{v}}_i$, when we have no \textit{a
priori} knowledge about the interference symbols ${\{s_k\}}_{k \ne
i}$ before performing symbol detection, and Fig.
\ref{fig:3D_GMM_after_iteration} represents the distribution of
${\bf{v}}_i$ after performing the PDA based detection, when we have
a relatively strong belief\footnote{In Fig.
\ref{fig:3D_GMM_after_iteration} the \textit{a priori} probability
vector ${\bf{P}}_n$ is set to $[0.1, 0.1, 0.7, 0.1]$, which is for
convenience of conceptual visualization. The actual maximum value of
the elements of the probability vector after performing the PDA
detection is typically near to $1.0$, which makes the other smaller
peaks corresponding to the less probable constellation symbols
almost vanish.} about the correct value of $s_i$ among all
legitimate constellation symbols $a_m$.

% \begin{figure}[tbp]
% \centering
%  %\includegraphics[width=2.8in, height = 2.4in]{Fig3.eps}
% \includegraphics[width=3.7in]{Fig6.eps}
% \caption{}
% \label{fig:GMM_PDF}
% \end{figure}

\section{The Proposed AB-Log-PDA Relying On A Priori Soft Feedback From the FEC Decoder}
\label{Sec:PDA_in_uncoded_system}
Based on the interference-plus-noise
distribution analysis of Section
\ref{Sec:interference_plus_noise_distribution_analysis}, below we
will elaborate on the proposed low-complexity AB-Log-PDA algorithm.
% capable of
% incorporating
% \footnote{Here ``outer source'' refers to
% any module which is independent of the PDA module and may provide
% some \textit{a priori} knowledge about the interfering symbols
% ${\{s_k\}}_{k \ne i}$. An example is the channel decoding module in
% the IDD scheme of Fig. \ref{fig:SI_PDA_schematic}.}
% the \textit{a
% priori} soft feedback about the interference symbols
% ${\{s_k\}}_{k \ne i}$ from the FEC decoder.
%
% Except presenting the different operations of the two algorithms, we
% will use ``the PDA'' to refer to either of them when presenting their common part without imposing ambiguity.
This algorithm uses the received signal $\bf{y}$, the channel
matrix $\bf{H}$, as well as the \textit{a priori} soft feedback gleaned from the FEC decoder, representing the soft estimates of the transmitted symbols ${\{s_i\}}_{i = 1, \cdots, N_t}$, as its input parameters, and generates the estimated decision probabilities for ${\{s_i\}}_{i = 1,
\cdots, N_t}$ as its output.
% For the sake of improving the numerical stability and
% accuracy, some of the key procedures are implemented in the log-domain.

As mentioned in Section \ref{Sec:interference_plus_noise_distribution_analysis},
% each symbol $s_i$ to be detected in
% Eq.~(\ref{eq:MIMO_Decoupled_Model}) is contaminated by both the
% noise $\bf{n}$ and the interfering signal ${\bf{u}}_i$. The noise is
% generally undesirable, but the interference might potentially become a
% useful signal, which is only not desired at some specific instant. Hence,
% if we can predict or estimate the interfering symbols
% $\{s_k\}_{k\ne i}$, the performance degradation imposed by
% ${\bf{u}}_i$ may be mitigated, or even completely eliminated under
% idealized conditions, such that a receive diversity order of $N_r$
% is approached.
if we can estimate the distribution of the interfering symbols $\{s_k\}_{k\ne i}$, the performance degradation imposed by
${\bf{u}}_i$ may be mitigated.
Although initially we do not have any \textit{a
priori} knowledge about the distribution of
$\{s_k\}_{k\ne i}$, we know exactly the distribution of the noise
$\bf{n}$ and we are also aware of the legitimate values of
$\{s_k\}_{k\ne i}$. Hence it is feasible to generate a coarse
estimate of ${\{s_i\}}_{i = 1,
\cdots, N_t}$ relying solely on the knowledge of the noise distribution and
the modulation constellation $\mathcal{A}$ at the
beginning. Based on these observations, we may assume that the
interference-plus-noise term ${\bf{v}}_i$ obeys a single $N_r$-variate
Gaussian distribution\footnote{Although this approximation is more
accurate when $N_t$ becomes larger, it is competent to produce a
coarse estimate of ${\{s_i\}}_{i = 1,
\cdots, N_t}$. }. In order to fully characterize the
complex random vector ${\bf{v}}_i$, which is not necessarily
\textit{proper}\footnote{The pseudo-covariance of a complex random vector $\bf{x}$ is defined as 
$ {\mathbb{C}}_p({\bf{x}}) \triangleq \mathbb{E}\left\{\left[{\bf{x}} - \mathbb{E}({\bf{x}})\right]\left[{\bf{x}} - \mathbb{E}({\bf{x}})\right]^T\right\}$. For a \textit{proper} complex random
variable, its pseudo-covariance vanishes, and it is sufficient to describe a
proper complex Gaussian distribution using only the mean and the
covariance\cite{Neeser:proper_Gaussian_process,Adali2011:complex_valued_SP}. However, for a coded system the in-phase and quadrature
components of the complex modulated signal $s_i$ might be
correlated, especially when the coding block-length is not long
enough. In this case, it is necessary to take into account an
additional second-order statistics, i.e. the pseudo-covariance
\cite{Neeser:proper_Gaussian_process},  to fully specify the
\textit{improper} complex Gaussian distribution in a generalized
manner. }, we specify the mean as
\begin{equation} \label{eq:vec_mean}
{\boldsymbol{\mu }}_i \triangleq \mathbb{E}({{\bf{v}}_i}) =  \sum\limits_{k \ne i} {\mathbb{E}(s_k) }
{\bf{h}}_k ,
\end{equation}
the covariance as
\begin{equation} \label{eq:vec_covariance}
{\boldsymbol{\Upsilon}}_i \triangleq \mathbb{C}({\bf{v}}_i) = \sum\limits_{k \ne i}
\mathbb{C}(s_k){\bf{h}}_k {\bf{h}}_k^H  + 2\sigma^2 {\bf{I}}_{N_r},
\end{equation}
and the pseudo-covariance as
\begin{equation} \label{eq:vec_pseudo-covariance}
{\underline{\boldsymbol{\Upsilon}}}_i \triangleq {\mathbb{C}_p}({\bf{v}}_i) = \sum\limits_{k \ne i}
\mathbb{C}_p(s_k) {\bf{h}}_k {\bf{h}}_k^T .
\end{equation}

We define an $(N_t \times M)$-element probability matrix
${\bf{P}}^{(z, z')}$, whose $(i,m)$-th element $P_m^{(z, z')} (s_i
|{\bf{y}})\triangleq P^{(z, z')}(s_i = a_m|{\bf{y}})$ is the
estimate of the probability that we have $s_i  = a_m $ at the
$(z,z')$-th iteration. More specifically, the integer $z'\ge 0$
denotes the inner iteration index of the AB-Log-PDA, while the
integer $z \ge 0$ is the index of the outer iteration between the
AB-Log-PDA and the soft FEC decoder, $i = 1, \cdots, N_t$ and
$m=1,\cdots, M$.
% For clarity,
% these probabilities are explicitly presented in form of
% the probability-matrix of Table \ref{table_PDA}.
Then, the $\mathbb{E}(s_k)$, ${\mathbb{C}}(s_k)$ and $\mathbb{C}_p(s_k)$ in (\ref{eq:vec_mean}), (\ref{eq:vec_covariance}) and (\ref{eq:vec_pseudo-covariance}) are given by
\begin{equation}\label{eq:symbol_mean}
\mathbb{E}(s_k)  = \sum\limits_{m = 1}^M {a_m P^{(z, z')} (s_k = a_m
|{\bf{y}})} ,
\end{equation}
\begin{equation}\label{eq:symbol_variance}
{\mathbb{C}}(s_k) = \sum\limits_{m = 1}^M {[a_m  - \mathbb{E} (s_k)
][a_m  - \mathbb{E}( s_k) ]^* P^{(z, z')}(s_k = a_m|{\bf{y}})} ,
\end{equation} and
\begin{equation}\label{eq:symbol_pseudo_variance}
\mathbb{C}_p(s_k) = \sum\limits_{m = 1}^M {[a_m  - \mathbb{E} (s_k)
]^2 P^{(z, z')}(s_k = a_m|{\bf{y}})},
\end{equation}
respectively.

Note that Eq. (\ref{eq:vec_mean}) - Eq.
(\ref{eq:symbol_pseudo_variance}) effectively use $(N_t -1)$ probability vectors
$\{{\bf{P}}^{(z, z')}(k, :)\}_{k\ne i}$ associated with the interfering
signal $\{s_k\}_{k\ne i}$  to model ${\bf{v}}_i$.  Since we do not
have any outer \textit{a priori} knowledge about the distribution of
$s_i|\bf{y}$ at the beginning, an all-zero LLR vector will be
provided as the input to the AB-Log-PDA, which is
equivalent to initializing $P^{(z, z')} (s_i = a_m |{\bf{y}})$ with a
uniform distribution, i.e.
\begin{equation}
 P^{(0, 0)}(s_i = a_m |{\bf{y}}) = \frac{1}{M},
\end{equation}
$\forall i = 1, \cdots, N_t$ and $ \forall m = 1,
\cdots, M$.

Based on the assumption that ${\bf{v}}_i$ obeys the Gaussian
distribution, ${\bf{y}}|s_i$ is also Gaussian distributed. Let us
now define
\begin{equation}\label{eq:Gaussian_variable}
{\bf{w}} \triangleq {\bf{y}}
- s_i {\bf{h}}_i  - \sum\limits_{k \ne i} {\mathbb{E} (s_k)
{\bf{h}}_k }
\end{equation}
and
\begin{equation}\label{eq:effective_metric}
\beta_{m, i}^{(z, z'+1)} \triangleq { - \left[
{\begin{array}{*{20}c}
   {\Re ({\bf{w}})}  \\
   {\Im ({\bf{w}})}  \\
\end{array}} \right]^T {\bf{\Lambda }}_i^{-1} \left[ {\begin{array}{*{20}c}
   {\Re ({\bf{w}})}  \\
   {\Im ({\bf{w}})}  \\
\end{array}} \right]},
\end{equation}
in which the \textit{composite} covariance matrix ${\bf{\Lambda }}_i$ is defined as \cite{Jia:CPDA}
\begin{equation} \label{eq:PDA_tot_var}
{\bf{\Lambda }}_i  \buildrel \Delta \over = \left[
{\begin{array}{*{20}c}
   {\Re \left( \boldsymbol{\Upsilon}_i + \underline{\boldsymbol{\Upsilon}}_i  \right)} & { - \Im \left( \boldsymbol{\Upsilon}_i - \underline{\boldsymbol{\Upsilon}}_i \right)}  \\
   {\Im \left( \boldsymbol{\Upsilon}_i + \underline{\boldsymbol{\Upsilon}}_i \right)} & {\Re \left( \boldsymbol{\Upsilon}_i - \underline{\boldsymbol{\Upsilon}}_i \right)}  \\
\end{array}} \right],
\end{equation}
where $\Re (\cdot)$ and $\Im (\cdot)$ represent the real and
imaginary part of a complex variable, respectively. Then the
likelihood function of ${\bf{y}}|s_i = a_m$ at the $(z, z'+1)$-st
iteration satisfies
\begin{equation} \label{eq:PDA_elem_prob}
p^{(z, z'+1)} ({\bf{y}}|s_i = a_m) \propto \exp\left(\beta_{m, i}^{(z, z'+1)}\right),
\end{equation}
where the symbol ``$\propto$'' means ``proportional to''. 

Upon invoking an approximate form of the Bayesian theorem\cite{Luo:PDA_Sync_CDMA, Pham:GPDA}, the
\textit{estimated} probability of symbol $s_i$ at the $(z, z'+1)$-st iteration
may be calculated as\footnote{$P^{(z, z')}(s_i =a_m)$ is ignored in (\ref{eq:PDA_app}), since it has been utilized for calculating the
likelihood of ${\bf{y}}|s_i =a_m$ at the $(z, z'+1)$-st iteration\ --- the same \textit{a priori} information should not be used multiple times in IDD scenarios.}
% \begin{eqnarray} \label{eq:PDA_app}
% P^{(z+1)}(s_i = a_m |{\bf{y}})  & \approx & \frac{{p^{(z+1)} ({\bf{y}}|s_i = a_m)}}{{\sum\limits_{m = 1}^M {p^{(z+1)} ({\bf{y}}|s_i = a_m)} }} \nonumber \\
% & = & \frac{\exp\left(\beta_{m,i}^{(z+1)} -\gamma\right)}{{\sum\limits_{m = 1}^M {\exp\left(\beta_{m,i}^{(z+1)} - \gamma\right)} }},
% \end{eqnarray}
\begin{eqnarray} \label{eq:PDA_app}
& & P^{(z, z'+1)}(s_i = a_m |{\bf{y}})  \nonumber \\
& \approx & \frac{{p^{(z, z'+1)} ({\bf{y}}|s_i = a_m)}}{{\sum\limits_{m = 1}^M {p^{(z, z'+1)} ({\bf{y}}|s_i = a_m)} }}  \nonumber \\
& = & \frac{\exp\left(\beta_{m,i}^{(z, z'+1)} -\gamma\right)}{{\sum\limits_{m = 1}^M {\exp\left(\beta_{m,i}^{(z, z'+1)} - \gamma\right)} }},
\end{eqnarray}
where $\gamma \triangleq \max\limits_{m =1, \cdots, M}{\beta_{m,i}^{(z, z'+1)}}$ is subtracted from $\beta_{m,i}^{(z, z'+1)}$ for enhancing the numerical stability.

\begin{table}[tbp]
\setlength{\tabcolsep}{1pt}
\renewcommand{\arraystretch}{1.0}
\extrarowheight 4pt \caption{Summary of the AB-Log-PDA based IDD algorithm}
\label{table:AB_Log_PDA} \centering
\begin{scriptsize}
\begin{tabular}{l}
\hlinewd{0.9pt}
%\hline
\textbf{Given} the received signal $\bf{y}$, the channel matrix $\bf{H}$ and the constellation $\mathcal{A}$. \\
\textbf{Step 1}. Set the initial values of the inner iteration index and outer iteration \\
index to $z' = 0$ and $z = 0$, respectively. Initialize the bit-based \textit{a priori}
LLRs \\
feedback from the FEC decoder as zeros. \\
% i.e. the symbol
%  probabilities are initialized as \\
% $P^{(z, z')}_m(s_i| {\bf{y}}) = 1/M$,
% $\forall {\rm{ }}i = 1,2, \cdots, N_t $ and $\forall {\rm{ }}m = 1,2,\cdots, M$. \\
\textbf{Step 2}.
Convert the \textit{a priori} LLRs feedback from the
FEC decoder to symbol \\
probabilities shown in probability matrix ${\bf{P}}^{(z,z')}$.  \\
\textbf{Step 3}. Using the values of $\left\{ {{\bf{P}}^{(z, z')}(k, :)}
\right\}_{k
\ne i} $, calculate $ P_m^{(z, z'+1)} (s_i|{\bf{y}})$ by \\
for $i = 1:N_t$ \\
\quad\quad calculate the statistics of the interference-plus-noise
term ${\bf{v}}_i$
using\\
\quad \quad (\ref{eq:vec_mean}) - (\ref{eq:symbol_pseudo_variance}),
as well as the inverse of ${\bf{\Lambda }}_i$ in (\ref{eq:PDA_tot_var}),\\
\quad\quad for $m = 1: M$ \\
\quad\quad\quad\quad calculate $ P_m^{(z, z'+1)} (s_i |{\bf{y}})$
using
(\ref{eq:Gaussian_variable}), (\ref{eq:effective_metric}), (\ref{eq:log_domain_PDA_app}) and (\ref{eq:prob_domain_PDA_app_final}).\\
\quad\quad end \\
end \\
\textbf{Step 4}. If $z'$ has reached a given number of inner iterations, go to Step 5. \\
Otherwise,
let $z' \leftarrow z' + 1$, and return to Step 3.
\\
\textbf{Step 5}. Convert the symbol probabilities $ P_m^{(z, z'+1)}
(s_i|{\bf{y}})$ to bit-based \\
LLRs, of which the extrinsic parts are delivered to the outer FEC decoder.\\
If $z$ has reach a given number of outer iterations, make hard decisions using the \\
 soft output of the FEC decoder. Otherwise, let $z \leftarrow z +1$, and return to Step 2.
\\

%\hline
\hlinewd{0.9pt}
\end{tabular}
\end{scriptsize}
\end{table}
As a further effort to improve the achievable numerical stability and
accuracy, the logarithmic-domain form of (\ref{eq:PDA_app}) is formulated as
% \begin{eqnarray}\label{eq:log_domain_PDA_app}
%  \psi_{m,i}^{(z+1)}& \triangleq & \ln\left(P^{(z+1)}(s_i = a_m |{\bf{y}})\right) \nonumber \\
% % & = & \ln\left(p^{(z+1)} ({\bf{y}}|s_i = a_m)\right) - \ln\left(\sum\limits_{m = 1}^M {p^{(z+1)} ({\bf{y}}|s_i = a_m)}\right) \nonumber \\
% & = & ({\tilde{\beta}}_{m,i}^{(z+1)}) - \ln\left(\sum\limits_{m = 1}^M {\exp\left({\tilde{\beta}}_{m,i}^{(z+1)} \right)}\right),
% \end{eqnarray}
\begin{eqnarray}\label{eq:log_domain_PDA_app}
 \psi_{m,i}^{(z, z'+1)} & \triangleq &  \ln\left[P^{(z, z'+1)}(s_i = a_m |{\bf{y}})\right] \nonumber \\
% & = & \ln\left(p^{(z+1)} ({\bf{y}}|s_i = a_m)\right) - \ln\left(\sum\limits_{m = 1}^M {p^{(z+1)} ({\bf{y}}|s_i = a_m)}\right) \nonumber \\
 & = & {\tilde{\beta}}_{m,i}^{(z, z'+1)} - \ln\left[\sum\limits_{m = 1}^M {\exp\left({\tilde{\beta}}_{m,i}^{(z, z'+1)} \right)}\right],
\end{eqnarray}
in which we have ${\tilde{\beta}}_{m,i}^{(z, z'+1)} \triangleq
\beta_{m,i}^{(z, z'+1)} -\gamma$, and the second term of the
right-hand-side expression may be computed by invoking the
``Jacobian logarithm'' of \cite{Hochwald:SD_near_capacity}. Alternatively, upon
employing the Max-log approximation, (\ref{eq:log_domain_PDA_app})
may be further simplified to
% \begin{equation} \label{eq:max_log_PDA_app}
%  \psi_{m,i}^{(z+1)} = {\tilde{\beta}}_{m,i}^{(z+1)} -  \max\limits_{m = 1, \cdots, M}{{\tilde{\beta}}_{m,i}^{(z+1)}}.
% \end{equation}
\begin{equation} \label{eq:max_log_PDA_app}
 \psi_{m,i}^{(z, z'+1)} \approx {\tilde{\beta}}_{m,i}^{(z, z'+1)}.
\end{equation}
As a result, the estimated decision probability of $s_i$ relying on (\ref{eq:log_domain_PDA_app}) and (\ref{eq:max_log_PDA_app}) is given by
\begin{equation} \label{eq:prob_domain_PDA_app_final}
P^{(z, z'+1)}(s_i = a_m |{\bf{y}})  \approx e^{ \psi_{m,i}^{(z, z'+1)}},
\end{equation}
which will update the value of $P^{(z, z')}(s_i = a_m |{\bf{y}})$ in the
probability matrix ${\bf{P}}^{(z, z')}$. Following the inner iterations within the AB-Log-PDA, \textit{if any}, the updated symbol probabilities
have to be
converted to the equivalent bit-based LLRs, whose \textit{extrinsic} constituent will be delivered to the outer FEC decoder of Fig.
\ref{fig:SI_PDA_schematic}. In turn, the extrinsic LLRs output by the
FEC decoder will be converted to symbol probabilities in the
next outer iteration, before feeding them into the AB-Log-PDA for generating new estimates of the
symbol probabilities.
For reasons of explicit clarity, the AB-Log-PDA algorithm relying
on the \textit{a priori} soft feedback generated by the FEC
decoder of Fig. \ref{fig:SI_PDA_schematic} is summarized in Table
\ref{table:AB_Log_PDA}.

\section{Extrinsic LLR Calculation for AB-Log-PDA in FEC-Coded MIMO Systems}
\label{Sec:PDA_in_coded_system}
% In a FEC-coded MIMO system employing
% the IDD scheme of Fig. \ref{fig:SI_PDA_schematic}, typically the
% extrinsic LLRs of the coded bits are exchanged between the SISO MIMO
% detector and the SISO FEC decoder. Therefore,
In order to integrate the
AB-Log-PDA into the IDD scheme, the AB-Log-PDA has to output
correct\footnote{As we will detail later, despite the fact that the  output symbol probability of the existing PDAs was typically interpreted as
the symbol APP,
it is actually not
the true APP, which ought to be proportional to both the likelihood and the \textit{a priori} probability\cite{Peter04:Bayesian_statistics}.} extrinsic LLRs for each of the FEC-coded bits, which is however,
not quite as straightforward as it seems at first sight, given that the output probabilities of the PDA
were interpreted as APPs in \cite{Luo:PDA_Sync_CDMA, Pham:GPDA, Liu:CPDA-apx, Jia:CPDA,
Shaoshi2011:B_PDA, Shaoshi2011:DPDA}.
%
% For the sake of clarity, in our ensuing exposition
We assume that
the components of the transmitted symbol-vector $\bf{s}$ are
obtained using the bit-to-symbol mapping function of $s_i =
\textrm{map}\left({{\bf{b}}_i}\right)$, $i=1,2,\cdots,N_t$, where
${\bf{b}}_i=\left[{b_{i,1},b_{i,2},\cdots, b_{il}, \cdots,
b_{i,M_b}}\right]^T $ $\in \left\{{+1,-1}\right\}^{M_b}$ is the
vector of bits mapped to symbol $s_i$. Additionally, we denote the
vector of bits corresponding to $\bf{s}$ as $\bf{b}$, which
satisfies $\bf{s}=\textrm{map}\left({\bf{b}}\right)$ and is formed
by concatenating the $N_t$ antennas' bit vectors ${\bf{b}}_1
,{\bf{b}}_2 , \cdots ,{\bf{b}}_{N_t }$, yielding ${\bf{b}} =  \left[
{{\bf{b}}_1^T ,{\bf{b}}_2^T , \cdots ,{\bf{b}}_{N_t }^T } \right]^T
= \left[ {b_1 ,b_2 , \cdots ,b_k , \cdots ,b_{M_b N_t } } \right]^T
\in \left\{ { + 1, - 1} \right\}^{M_b N_t } $. Hence the indices of
$b_{il}$ and $b_k$ are related to each other by $k = {M_b}(i-1)+l$.

Note that the AB-Log-PDA algorithm finally outputs the
\textit{estimated} symbol probabilities of $P(s_i = a_m|{\bf{y}})$, where the iteration indices are omitted without
causing any confusion. Additionally, it provides the likelihood functions of
$p({\bf{y}}|s_i= a_m)$ conditioned on $s_i$ as its intermediate
output. By contrast, the classic candidate-search based approach
outputs the likelihood function of $p({\bf{y}}|{\bf{b}})$ [or
equivalently, $p({\bf{y}}|{\bf{s}})$] conditioned on the bit vector
$\bf{b}$ (or symbol vector $\bf{s}$), and calculates the bit-based
extrinsic LLRs by using $p({\bf{y}}|{\bf{b}})$ [or
$p({\bf{y}}|{\bf{s}})$]. Below we will demonstrate that the candidate-search based
approach of computing the bit-based extrinsic LLRs is not feasible for
the AB-Log-PDA algorithm. In other words, we cannot obtain $p({\bf{y}}|{\bf{b}})$ or
$p({\bf{y}}|{\bf{s}})$ based on $P(s_i = a_m|{\bf{y}})$ and/or $p({\bf{y}}|s_i= a_m)$. Instead, we will demonstrate that there exists a
simpler method of directly obtaining the bit-based extrinsic LLRs based on the
output of the AB-Log-PDA.
\subsection{Challenges in Calculating Extrinsic LLRs for the PDA Based Methods}
In principle, for a MIMO system characterized by Eq.
(\ref{eq:MIMO_model}),
the classic approach of deriving bit-based extrinsic LLRs is
based on the likelihood function of $p({\bf{y}}|{\bf{b}})$ or
$p({\bf{y}}|{\bf{s}})$. Specifically, 
% the \textit{a posteriori}
the extrinsic
LLR of $b_{il}$ (or $b_{k}$) is given by \cite{Hochwald:SD_near_capacity}
\begin{equation}\label{eq:L_E_raw}
  L_E(b_{il}|{\bf{y}}) = L_E(b_{k}|{\bf{y}})   =  \ln\frac{\sum\limits_{\forall {\bf{b}}\in \mathbb{B}_{k}^+}p({\bf{y}}|{\bf{b}})\exp (\frac{1}{2}{\bf{b}}_{[k]}^T{\bf{L}}_{A,[k]}) }
{\sum\limits_{\forall {\bf{b}}\in \mathbb{B}_{k}^-}p({\bf{y}}|{\bf{b}})\exp (\frac{1}{2}{\bf{b}}_{[k]}^T {\bf{L}}_{A,[k]}) },
\end{equation}
where $\mathbb{B}_{k}^\pm$ denotes the set of $2^{N_tM_b-1}$ legitimate bit vectors $\bf{b}$
having $b_{k} = \pm 1$, and ${\bf{b}}_{[k]} = [b_{1}, \cdots, $ $b_{j}, \cdots,
 b_{N_tM_b} ]_{j\ne k}^T$ represents a truncated version of $\bf{b}$
 excluding $b_k$, while ${\bf{L}}_{A,[k]} $ represents the \textit{a priori} LLRs
corresponding to ${\bf{b}}_{[k]}$. Eq. (\ref{eq:L_E_raw}) indicates
that $ L_E(b_{k}|{\bf{y}})$ is determined by $p({\bf{y}}|{\bf{b}})$,  and by the \textit{a priori} LLRs of the
other bits conveyed by a single symbol vector $\bf{s}$.
However, below we will prove that it is infeasible to invoke this
approach to calculate bit-based extrinsic LLRs for the family of
PDA based algorithms including the AB-Log-PDA.
\begin{proposition}\label{proposition:non_applicable_theorem}
For all PDA algorithms which output the probabilities
$P(s_i|{\bf{y}})$, or the likelihood functions $p({\bf{y}}|s_i)$, the
bit-based extrinsic LLR $L_E(b_{k}|{\bf{y}})$ cannot be calculated using the candidate-search method which relies on $p({\bf{y}}|\bf{b})$ or
$p({\bf{y}}|\bf{s})$.
\end{proposition}
\begin{IEEEproof}
% This theorem is readily proved by invoking Lemma
% \ref{lem:non_reducible_lemma}.
Define a non-zero random vector
${\bf{s}} = [s_1, s_2, \cdots, s_{N_t}]^T$, and a non-zero random
vector $\bf{y}$, where $s_i$ and $s_j$ are independent of each other
in the absence of \textit{a priori} knowledge, $i \ne j$, $i, j = 1, \cdots, N_t$, and assume that
$\bf{y}$ is associated with $\bf{s}$ by the function of ${\bf{y}} =
f({\bf{s}})$. We have
\begin{eqnarray}
P({\bf{y}}|{\bf{s}}) & = & P({\bf{y}}|s_1, s_2, \cdots, s_{N_t}) \nonumber \\
& = & \frac{P({\bf{y}}, s_1, s_2, \cdots, s_{N_t})}{P(s_1, s_2,
\cdots, s_{N_t})} \nonumber \\
% & = & \frac{P(s_1, s_2, \cdots,
% s_{N_t}|{\bf{y}})P({\bf{y}})}{P(s_1)P(s_2)\cdots P(s_{N_t})}
% \nonumber \\
& = &
\frac{P(s_1|{\bf{y}})P(s_2,\cdots,s_{N_t}|{\bf{y}},s_1)P({\bf{y}})}{P(s_1)P(s_2)\cdots
P(s_{N_t})},
% & = & \frac{P(s_1|{\bf{y}})P(s_2|{\bf{y}}, s_1)\cdots
%P(s_{N_t}|{\bf{y}},s_1, s_2, \cdots,
%s_{N_t})P({\bf{y}})}{P(s_1)P(s_2)\cdots P(s_{N_t})}
\end{eqnarray}
where $P(s_2,\cdots,s_{N_t}|{\bf{y}},s_1)$ can be further expanded
as
\begin{equation}\label{eq:prob_factorization}
P(s_2|{\bf{y}}, s_1)P(s_3|{\bf{y}},s_1, s_2)\cdots
P(s_{N_t}|{\bf{y}},s_1, s_2, \cdots, s_{N_t-1}).
\end{equation}
Note that the conditions $s_i$ associated with each single
probability in (\ref{eq:prob_factorization}) cannot be removed, 
which implies that it is infeasible to further simplify each
probability in (\ref{eq:prob_factorization}). 
In other words, we have $P({\bf{s}}|{\bf{y}}) \ne P(s_1|{\bf{y}})P(s_2|{\bf{y}}) \cdots P(s_{N_t}|{\bf{y}})$, which implies that in a converging connection of the acyclic, directed graph representation of Bayesian Networks, the presence of knowledge as regards to the child-node makes the parent-nodes conditionally dependent. Again, this is a standard result in Bayesian Networks \cite{Pearl_1988_prob_reasoning}.
Therefore, the probability $P({\bf{y}}|{\bf{s}})$ cannot be
exactly expressed as a function of any probabilities of $P(s_i)$, $P(\bf{y})$,
$P({\bf{y}}|s_i)$ and/or $P(s_i|{\bf{y}})$. Hence the proof of
Proposition \ref{proposition:non_applicable_theorem} is established.
\end{IEEEproof}

% Then by multiplying the numerator and denominator with $\exp[-\frac{1}{2}\sum\nolimits_{j=0, j\ne l}^{M_b-1}L_A(b_{il})]$,
% (\ref{eq:L_E_raw}) may be rewritten as
% \begin{equation}
% L_E(b_{il}) = \ln\frac{\sum\limits_{\forall a_m\in \mathcal{A}_l^+}P({\bf{y}}|s_i = a_m)\exp (\frac{1}{2}{\bf{b}}_{[l]}^T{\bf{L}}_{A,[l]}) }
% {\sum\limits_{\forall a_m\in \mathcal{A}_l^-}P({\bf{y}}|s_i = a_m)\exp (\frac{1}{2}{\bf{b}}_{[l]}^T {\bf{L}}_{A,[l]}) }
% \end{equation}

\subsection{Calculating Extrinsic LLRs for AB-Log-PDA}
Due to Proposition \ref{proposition:non_applicable_theorem}, the
candidate-search based approach of calculating bit-based extrinsic  LLRs
is not applicable to the family of PDA algorithms. Let $\mathcal{A}_{l}^{\pm}$ denote the set of $M/2$ constellation points whose $l$-th bit is $\pm 1$. Then, alternatively, the extrinsic LLR of $b_{il}$ may be
rewritten as\footnote{The relationship of $P({\bf{y}}|b_{il} = \pm
1) = \sum\limits_{\forall a_m\in \mathbb{A}_{l}^{\pm}}
P({\bf{y}}|s_i = a_m)P(s_i = a_m) $ holds only for single-antenna
systems. One may argue nonetheless that it also seems to make sense
for multiple-antenna systems, because the value of $b_{il}$ is
directly determined by the value of the symbol $s_i$ at the $i$-th
antenna, rather than by the values of other symbols $s_j$, $j\ne i$.
This line of argument is however, deceptive for the MIMO scenario
considered. The rationale is that $\bf{y}$ is associated with the
symbol vector $\bf{s}$ (or bit vector $\bf{b}$), rather than only
with the specific symbol of any specific antenna.}
% \begin{eqnarray}
%   L_D(b_{il}|{\bf{y}}) & = & \ln\frac{P(b_{il} = +1|{\bf{y}})}{P(b_{il} = -1|{\bf{y}})} \nonumber \\
%   & = & \ln\frac{\sum\limits_{\forall a_m\in \mathcal{A}_l^+}P(s_i = a_m|{\bf{y}})}{\sum\limits_{\forall a_m\in \mathcal{A}_l^-}P(s_i = a_m|{\bf{y}})} \nonumber \\
% & = & L_E(b_{il}|{\bf{y}}) + \underbrace{\ln\frac{P(b_{il} =
% +1)}{P(b_{il} = -1)}}_{L_A(b_{il})}.
% \end{eqnarray}
% \begin{equation}
%   L_D(b_{il}|{\bf{y}})
% % =  \ln\frac{P(b_{il} = +1|{\bf{y}})}{P(b_{il} = -1|{\bf{y}})}
%    =  \ln\frac{\sum\limits_{\forall a_m\in \mathcal{A}_l^+}P(s_i = a_m|{\bf{y}})}{\sum\limits_{\forall a_m\in \mathcal{A}_l^-}P(s_i = a_m|{\bf{y}})}
%  =  L_E(b_{il}|{\bf{y}}) + \underbrace{\ln\frac{P(b_{il} =
% +1)}{P(b_{il} = -1)}}_{L_A(b_{il})}.
% \end{equation}
% In other words, we have
% \begin{eqnarray}\label{eq:symbol_based_L_e}
%   L_E(b_{il}|{\bf{y}}) & = & \ln\frac{\sum\limits_{\forall a_m\in \mathcal{A}_l^+}P(s_i = a_m|{\bf{y}})}{\sum\limits_{\forall a_m\in \mathcal{A}_l^-}P(s_i = a_m|{\bf{y}})} \nonumber \\
% & & - \underbrace{\ln\frac{P(b_{il} = +1)}{P(b_{il} =
% -1)}}_{L_A(b_{il})}.
% \end{eqnarray}
\begin{equation}\label{eq:symbol_based_L_e}
  L_E(b_{il}|{\bf{y}})  =  \underbrace{\ln\frac{\sum\limits_{\forall a_m\in \mathcal{A}_l^+}P(s_i = a_m|{\bf{y}})}{\sum\limits_{\forall a_m\in \mathcal{A}_l^-}P(s_i = a_m|{\bf{y}})}}_{L_D(b_{il}|{\bf{y}})}
  - \underbrace{\ln\frac{P(b_{il} = +1)}{P(b_{il} =
-1)}}_{L_A(b_{il})}, 
\end{equation}
where $L_D(b_{il}|{\bf{y}})$ and $L_A(b_{il})$ denote the \textit{a posteriori} and
\textit{a priori} LLRs of $b_{il}$, respectively. 
It is noteworthy that (\ref{eq:symbol_based_L_e}) represents a
simple approach of generating the bit-based extrinsic LLR of
$L_E(b_{il}|{\bf{y}})$, as long as the \textit{true} symbol APP of
$P(s_i = a_m|{\bf{y}})$ can be obtained.

However, although we can directly obtain the estimated symbol
probabilities of $P(s_i = a_m|{\bf{y}})$ from the output of the AB-Log-PDA,
as shown in (\ref{eq:PDA_app}), our study shows that this sort of
estimated symbol probabilities, interpreted as symbol APPs in \cite{Luo:PDA_Sync_CDMA, Pham:GPDA, Liu:CPDA-apx, Jia:CPDA,
Shaoshi2011:B_PDA, Shaoshi2011:DPDA}, fail to generate the correct
bit-based extrinsic  LLRs, when invoking
(\ref{eq:symbol_based_L_e})\footnote{In fact, if
$L_E(b_{il}|{\bf{y}})$ is calculated by substituting the estimated
symbol probabilities of $P(s_i = a_m|{\bf{y}})$, i.e. the output of the
AB-Log-PDA, into Eq. (\ref{eq:symbol_based_L_e}), the slope of the resultant BER
curve of the IDD scheme of Fig. \ref{fig:SI_PDA_schematic} remains almost horizontal upon increasing signal-to-noise ratio (SNR) values. Due to
the limitations of space, this flawed BER curve
is not presented in this paper.}.
%because they are calculated using an
%approximated form of the Bayesian theorem, rather than the exact Bayesian theorem. To elaborate a little
%further, let us revisit (\ref{eq:PDA_app}). Rigorously, the
%symbol APPs are supposed to be calculated by
%\begin{equation} \label{eq:PDA_app_true}
%P(s_i = a_m |{\bf{y}}) = \frac{{p ({\bf{y}}|s_i = a_m)}P(s_i =
%a_m)}{{\sum\limits_{m = 1}^M {p({\bf{y}}|s_i = a_m)}P(s_i=a_m) }}.
%\end{equation}
%Therefore, (\ref{eq:PDA_app}) actually holds only for the initial iteration
%when the \textit{a priori} knowledge is not available and hence the
%initial values of all the symbol probabilities are assumed to be
%$P(s_i = a_m) = \frac{1}{M}$.
Therefore, the results of (\ref{eq:PDA_app}) should not be interpreted as symbol APPs satisfying (\ref{eq:symbol_based_L_e}),
but rather as the normalized symbol likelihoods.
Based on this insight, the bit-based extrinsic
LLRs of the AB-Log-PDA may be obtained by directly employing the approximate Bayesian Theorem based symbol
probabilities of (\ref{eq:PDA_app}) as follows.
\begin{conjecture} \label{conjecture:L_e_AB_Log_PDA}
The bit-based extrinsic LLR of the AB-Log-PDA algorithm relying on
(\ref{eq:PDA_app}) is given by
\begin{equation}\label{eq:pseudo_symbol_APP_based_L_e}
 L_E(b_{il}|{\bf{y}}) \approx \ln\frac{\sum\limits_{\forall a_m\in
\mathcal{A}_l^+}P(s_i = a_m|{\bf{y}})}{\sum\limits_{\forall a_m\in
\mathcal{A}_l^-}P(s_i =
  a_m|{\bf{y}})},
\end{equation}
where $P(s_i = a_m|{\bf{y}})$ is calculated by invoking
(\ref{eq:prob_domain_PDA_app_final}).
\end{conjecture}
%\begin{IEEEproof}
%The proof can be made based on Theorem
%\ref{theorem:non_equivalent_theorem}.
%\end{IEEEproof}

The $L_E(b_{il}|{\bf{y}}) $ values calculated from
(\ref{eq:pseudo_symbol_APP_based_L_e}) using the normalized symbol likelihoods are typically not equivalent to $L_E(b_{il}|{\bf{y}}) $
calculated from (\ref{eq:symbol_based_L_e}) using the true symbol
APPs, but nonetheless, they constitute a good approximation of the
latter without inducing any significant performance loss, as it will
be demonstrated by our simulations in Section \ref{Sec:Simulations}.
As a result, the classic IDD receiver structure is simplified, as shown in Fig.
\ref{fig:SI_PDA_schematic}, where we have ${\bf{L}}_{E_2} =
{\bf{L}}_{D_2}$, rather than ${\bf{L}}_{E_2} = {\bf{L}}_{D_2} -
{\bf{L}}_{A_2}$.

\section{Simulation Results and Discussions}
\label{Sec:Simulations}
% \subsection{Original PDA with fixed inner iteration}
% \subsection{Modified PDA without inner iteration}
% \subsection{Modified PDA with fixed inner iteration}
% \subsection{Modified PDA with threshold rather than a fixed inner iteration}
% \subsection{numerical stability vs. implementation}
In this section, the performance of the proposed AB-Log-PDA based
IDD scheme is characterized with the aid of both semi-analytical
extrinsic information transfer (EXIT) charts and Monte-Carlo
simulations. Additionally, the complexity of the proposed AB-Log-PDA
based IDD scheme is analyzed, which further confirms the attractive
performance versus complexity tradeoff achieved by the proposed AB-Log-PDA
based IDD scheme. The FEC employed is the parallel concatenated recursive systematic convolutional (RSC) code based turbo code
having a coding rate\footnote{As usual, half of the parity bits generated by each of the two RSC codes are punctured.} of $R = \frac{k}{n} = 1/2$, constraint
length of $L = 3$ and generator polynomials of $(7, 5)$ in octal form. The turbo code is decoded by
the Approximate-Log-MAP algorithm using $it_{tc} = 4$ inner iterations. The interleaver employed is the $2400$-bit random sequence
interleaver. The remaining scenario-dependent simulation parameters are shown in the respective figures, where the MIMO arrangement is represented
in form of
$(N_t \times N_r)$.
%  Table \ref{Table:simu_parameter}.
\subsection{Performance of the AB-Log-PDA based IDD}
\textit{1) Impact of inner PDA iterations} 

\begin{figure}[tbp]
\centering
\includegraphics[width=3.6in]{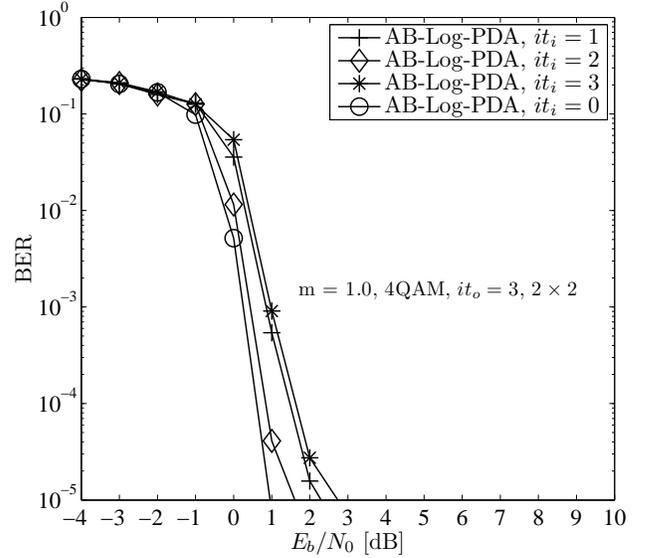}
\caption{Impact of the number of inner iterations on the achievable
BER of the AB-Log-PDA based IDD schemes.}
\label{fig:inner_iteration_on_BER}
\end{figure}
In Fig. \ref{fig:inner_iteration_on_BER}, we investigate the impact of the number of inner iterations
within the AB-Log-PDA algorithm on the achievable performance of the IDD scheme, which
is degraded upon increasing the number of inner iterations of the AB-Log-PDA, despite the fact that the computational complexity increases
dramatically. This implies that the optimal number of
inner iterations of the AB-Log-PDA conceived for the IDD receiver is $it_i = z' = 0$. 
It should be noted that for other types of iterative receivers, the inner iterations often refer to the iterations within the FEC-decoder, in which  
% where the FEC-decoder is not the inner, but the outer component of a serially concatenated scheme. 
% In these iterative FEC-decoders, 
typically the MAP algorithm and its variants are employed. In that context, increasing the number of inner iterations typically improves the iterative receiver's performance, which is in contrast to the impact of the inner PDA detector's iterations, as shown in Fig. \ref{fig:inner_iteration_on_BER}. 

The reasons as to why the inner PDA iterations fail to provide BER improvement can be understood from three different perspectives, as detailed below. 

i) The PDA method is reconfigurable, and both the inner PDA iterations as well as the outer iterations play a similar \textit{role} with respect to the PDA detector module in our IDD scenario, but the soft information provided by the two sorts of iterations has a different \textit{quality}. Firstly, when the number of IAI components is insufficiently high for the central limit theorem to prevail, there is an inevitable Gaussian approximation error, even if the PDA method has converged to its best possible estimate. This approximation error is more severe, when the soft information provided by the Gaussian approximation in each inner PDA iteration is unreliable, because error propagation will occur during the process of inner PDA iterations. Furthermore, if we look at the PDA detector module in isolation, the inner PDA iterations and the outer iterations play a similar \textit{role} -- both of them are responsible for providing the input soft information for the next round of Gaussian approximation. This Gaussian approximation  procedure is identical for the two sorts of iterations, while the \textit{quality} of the soft information provided by the two types of iterations is different. Additionally, compared to the scenario of uncoded systems, where the PDA method can only rely on its own knowledge of the transmitted/received signal and its own inner iterations, in FEC-coded systems the Gaussian approximation error can be mitigated more effectively by the improved-reliability soft information fed back by the FEC decoder. 
% In other words, the improved-reliability soft information input provided by the outer FEC decoder is more beneficial for the PDA method than the less reliable soft information input generated by the PDA method relying entirely on its own knowledge. 
Therefore, the inner PDA iterations can be replaced by the more efficient outer iterations in the IDD scenario considered. 

\begin{figure}[tbp]
\centering
\includegraphics[width=3.6in]{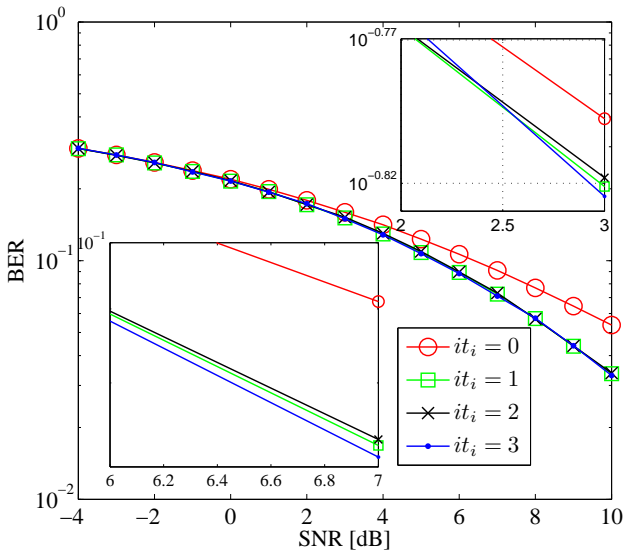}
\caption{Observation of the fine details of the impact of inner PDA iterations on the achievable performance of a PDA-based detector in an uncoded MIMO system, where we have $N_t = N_r = 2$, and $4$QAM is used.}
\label{Fig_reviewer_2:impact_ite_PDA_on_BER_only_data}
\end{figure}
ii) The convergence profile of the PDA method is not monotonic. In engineering/optimization problems two typical types of convergence behaviors may be observed for a function/sequence. Namely, the function/sequence may monotonically approach its optimum, or may fluctuate during the process of approaching its optimum --- hopefully without getting trapped in a local optimum. 
Upon observing Fig. \ref{Fig_reviewer_2:impact_ite_PDA_on_BER_only_data} as to the fine details of the impact of inner PDA iterations on the achievable performance of the PDA detector in an uncoded MIMO system, we find that the convergence behavior of the PDA method belongs to the second type. This particular convergence behavior of the PDA has not been reported in the open literature, because in uncoded systems hard decisions are made based on the output symbol probabilities of the PDA method. Hence the resultant BER performance fluctuation may remain so trivial that it may be regarded as being unchanged after several inner PDA iterations. However, as we can see from Fig. \ref{Fig_reviewer_2:impact_ite_PDA_on_BER_only_data}, the BER performance of $it_i = 1, 2, 3$ actually exhibits some degree of fluctuations. 

\begin{figure}[tbp]
\centering
\includegraphics[width=3.6in]{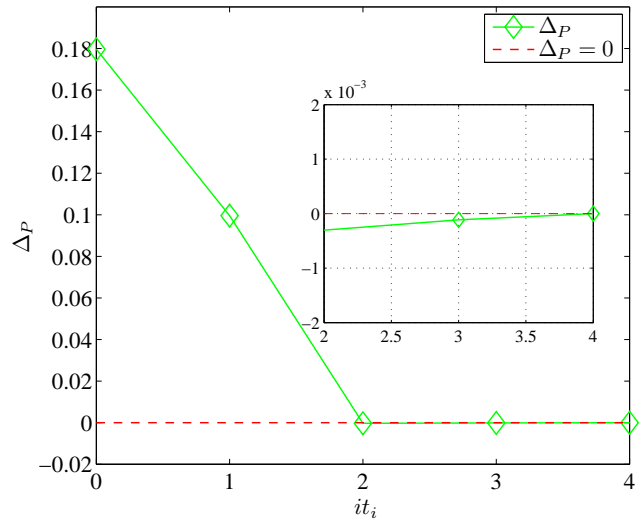}
\caption{Observation of the fine details of the probability convergence profile: the probability gap between the previous iteration and the current iteration of a PDA-based detector when we look at a single symbol probability $P(s_i = a_m|{\bf{y}})$ in an uncoded MIMO systems, where we have $N_t = N_r = 2$, $4$QAM.}
\label{Fig_reviewer_2:single_prob_dif_ite_zoom}
\end{figure}
These fluctuations can be further confirmed by tracking the changes of a single symbol's probability $P(s_i = a_m |{\bf{y}})$ during the inner PDA iterations, as shown in Fig. \ref{Fig_reviewer_2:single_prob_dif_ite_zoom}.
In uncoded systems, the PDA method is regarded to be converged when the probability changes obey $\Delta P = |P(\textrm{current iteration}) - P(\textrm{previous iteration})| < \epsilon$, where the threshold $\epsilon$ is a small positive real number. In Fig. \ref{Fig_reviewer_2:single_prob_dif_ite_zoom} we show how $\Delta P$ changes upon increasing the number of inner PDA iterations in the context of a $4$QAM aided uncoded $(2\times 2)$-element MIMO system, where we have $\epsilon = 0.001$. Again, although a superficial observation shows that $\Delta P$ remains more or less unchanged after two iterations, $\Delta P$ actually exhibits a modest fluctuation, because we have both positive and negative values of $\Delta P$ during the iterations. 

However, the trivial fluctuation at the soft output of the PDA detector module may induce an augmented BER fluctuation at the output of the FEC-coded system. Let us consider for example a symbol probability vector of ${\bf{p}}_1 = [0.35, 0.58, 0.06, 0.01]$, which represents our belief as regards to $s_i = a_1, a_2, a_3, a_4$, respectively. In uncoded systems, what we care about is, which specific probability is the maximum. In this case we will choose $s_i = a_2$, and a modest fluctuation from ${\bf{p}}_1$ to ${\bf{p}}_2 = [0.38, 0.53, 0.06, 0.01]$ will not lead to a different decision. However, in FEC-coded systems, we care about both the amplitude and the sign of the LLRs. A modest fluctuation in the probability vector ${\bf{p}}_1$ may alter some of the resultant LLRs that are near zero, so that they fluctuate between positive/negative values and hence might cause more severe decision errors. 

\begin{figure}[tbp]
\centering
\includegraphics[width=3.6in]{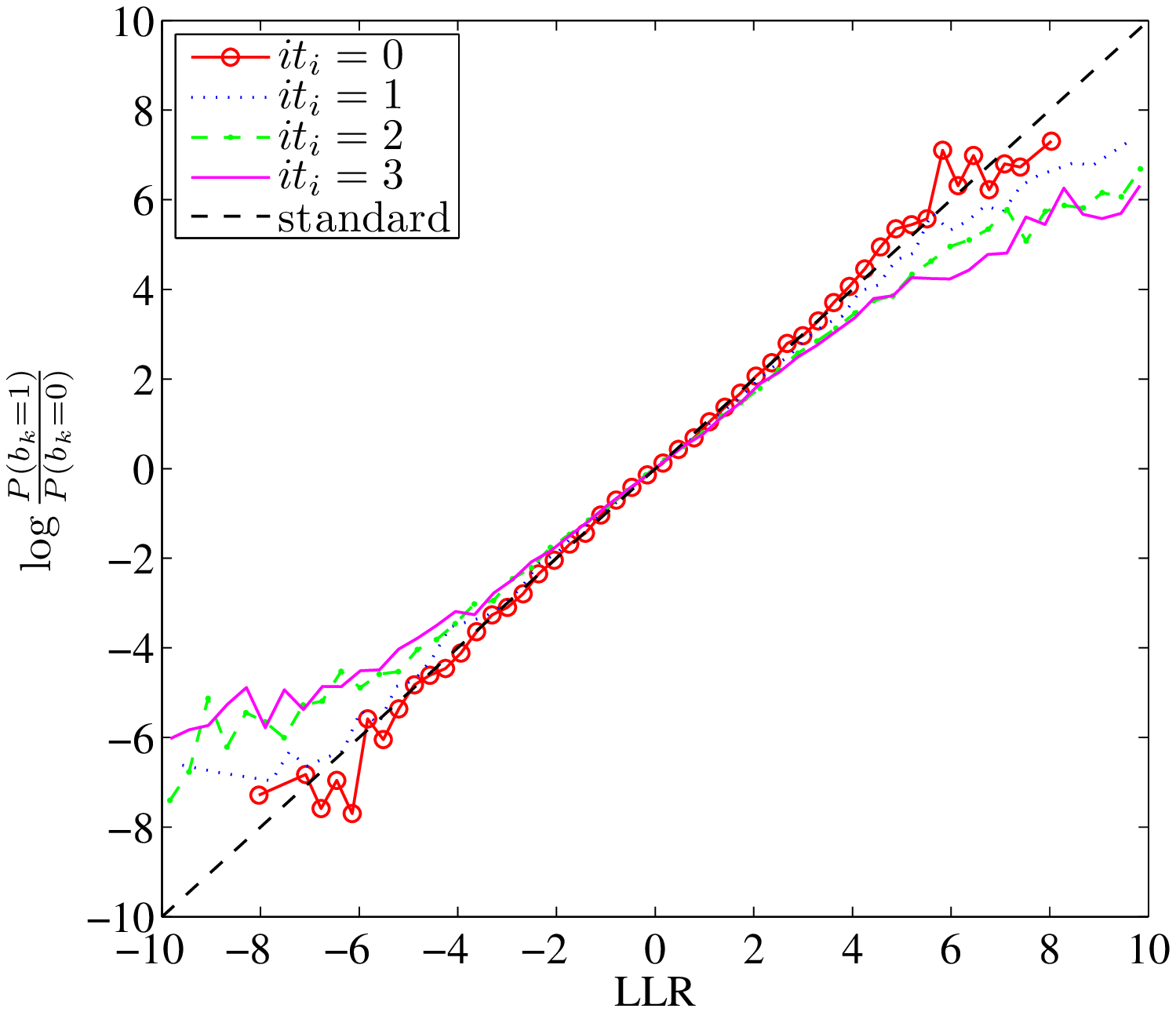}
\caption{The impact of inner PDA iterations on the consistency of LLRs output by the AB-Log-PDA detector.}
\label{Fig:consistency_test_AB-Log-PDA}
\end{figure}
iii) The inner PDA iterations degrade the quality of the LLRs output by the AB-Log-PDA.    
In Fig. \ref{Fig:consistency_test_AB-Log-PDA} we show the impact of the inner PDA iterations on the quality of the LLRs at the output of the AB-Log-PDA by testing the so-called consistency condition \cite{hagenauer2004:EXIT_chart} of these LLRs. As seen from Fig. \ref{Fig:consistency_test_AB-Log-PDA}, the consistency profile of the LLRs at the output of the AB-Log-PDA is degraded upon increasing the number of inner PDA iterations. This observation provides another perspective, confirming that it may in fact be detrimental to include inner PDA iterations, when a PDA-based IDD receiver is considered in FEC-coded systems. Therefore,
we dispense with inner iterations in the AB-Log-PDA and set $it_i = 0$ in our forthcoming simulations.

\textit{2) Impact of outer iterations}
 
\begin{figure}[tbp]
\centering
\includegraphics[width=3.7in]{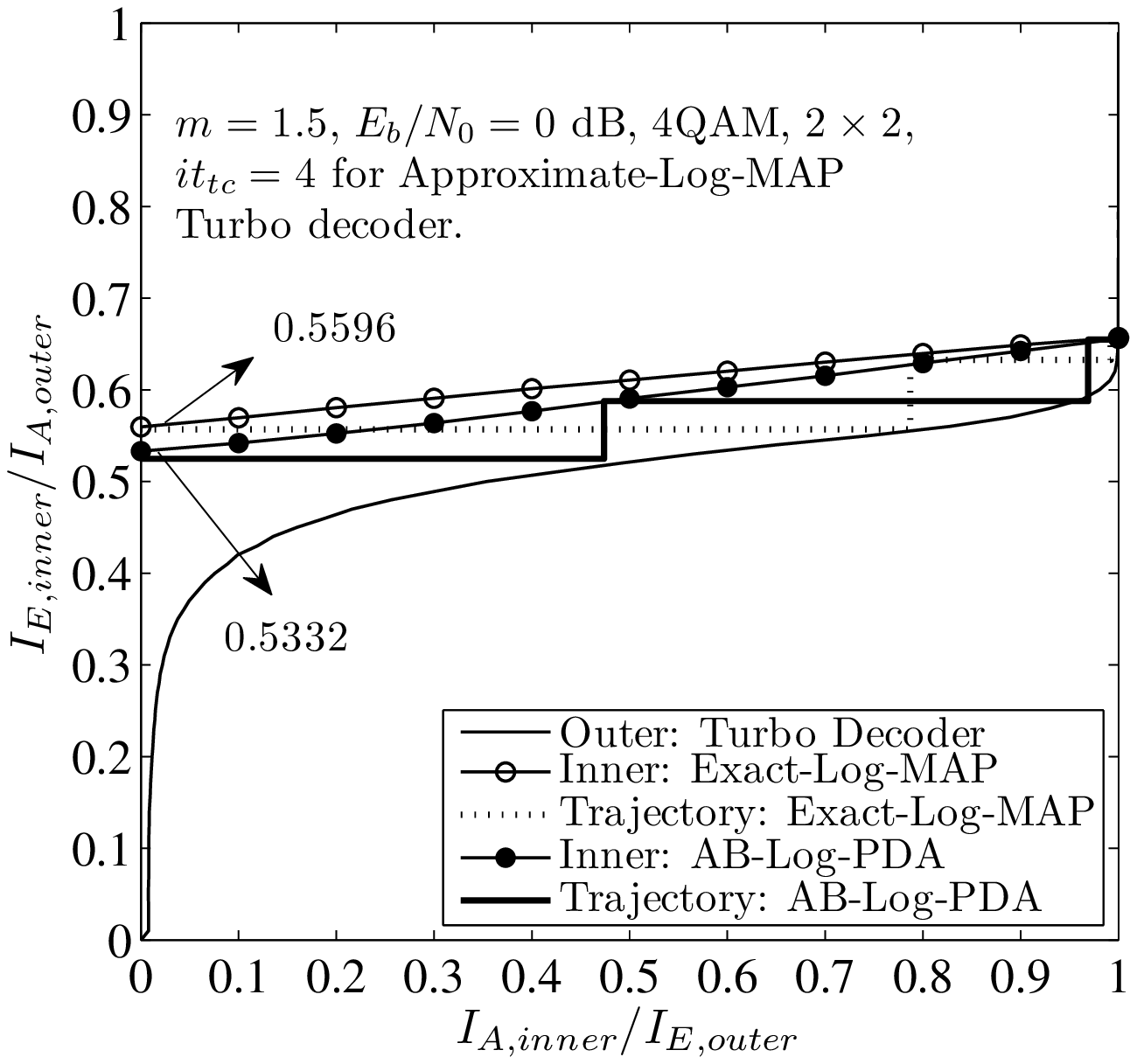}
\caption{EXIT chart analysis of the AB-Log-PDA ($it_i = 0$) and the Exact-Log-MAP based
IDD schemes.} \label{fig:EXIT_Trajectory_comparison}
\end{figure}

Fig. \ref{fig:EXIT_Trajectory_comparison} compares the convergence
behavior of the proposed AB-Log-PDA based IDD with $it_i = 0$ and that of the
optimal Exact-Log-MAP
% \footnote{The look-up table based
% Approximate-Log-MAP detector and the Max-Log-MAP detector are not
% considered here because they will induce performance-loss to some
% extent, though often very small.}
based IDD scheme using EXIT chart \cite{ten_brink2001:EXIT_chart}
analysis, where the EXIT
curve of the AB-Log-PDA is close to that of the Exact-Log-MAP. For
example, when the \textit{a priori} mutual information is
$I_{A,inner} = 0$, the extrinsic mutual information of the
AB-Log-PDA and of the Exact-Log-MAP is $I_{E, outer} = 0.5332 $ and
$I_{E, outer} = 0.5596$, respectively. This indicates that the
performance of the AB-Log-PDA is close to that of the Exact-Log-MAP
in the scenario considered. Additionally, the detection/decoding
trajectories indicate that both the AB-Log-PDA and the Exact-Log-MAP
based IDD schemes converge after three iterations, although the
respective performance improvements achieved at each iteration are
different.

\begin{figure}[t]
\centering
\includegraphics[width=3.6in]{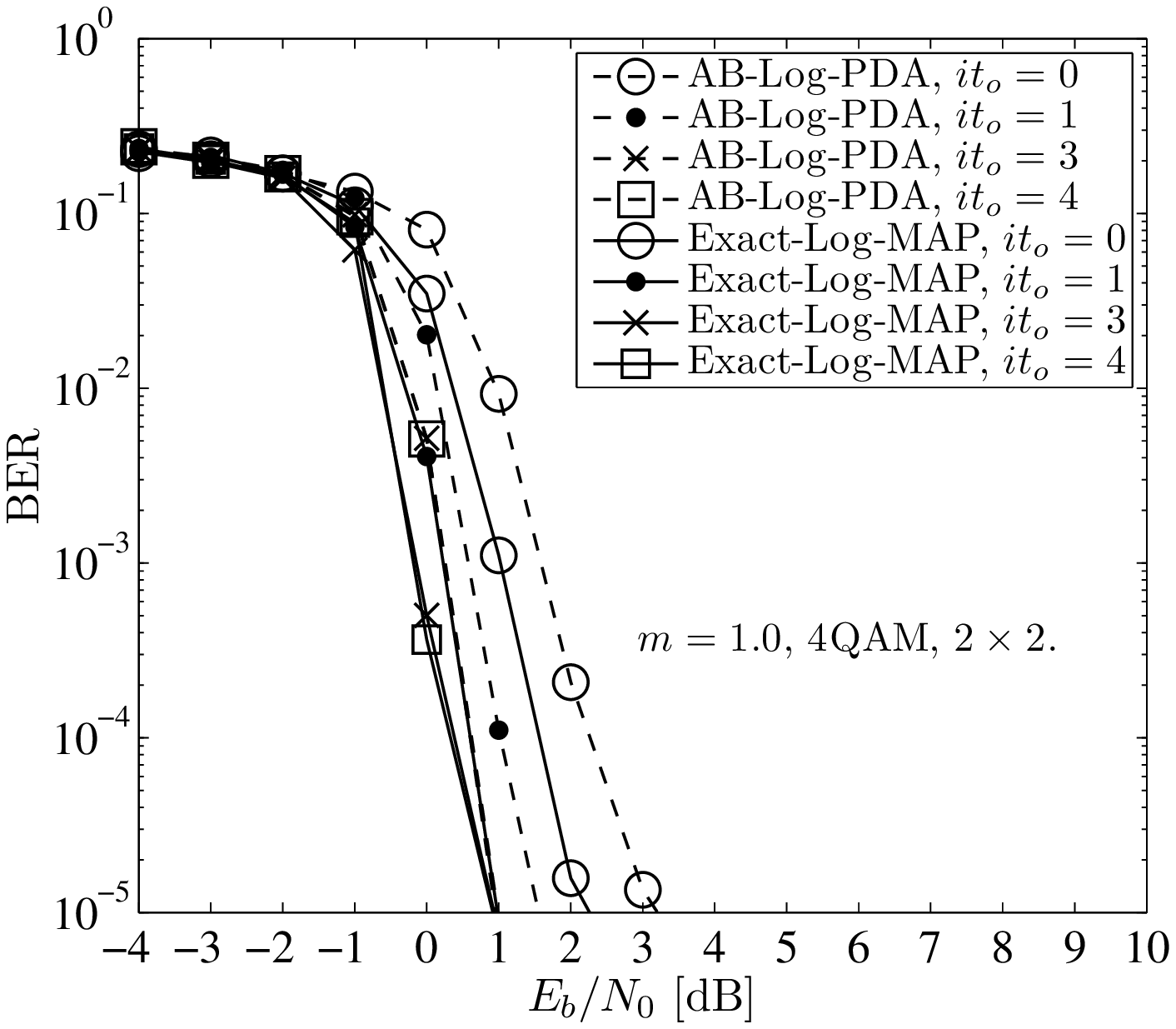}
\caption{Impact of the number of outer iterations on the achievable
BER of the AB-Log-PDA ($it_i = 0$) and the Exact-Log-MAP based IDD schemes.}
\label{fig:outer_iteration_on_BER}
\end{figure}
The above EXIT chart based  performance prediction and the
convergence behavior of the IDD schemes considered are also
characterized by the BER performance results of Fig.
\ref{fig:outer_iteration_on_BER}, where the Nakagami-$m$ fading
parameter is set to $m = 1.0$, which corresponds to the Rayleigh
fading channel. Observe from Fig. \ref{fig:outer_iteration_on_BER}
that the performance of the AB-Log-PDA based IDD scheme is improved
upon increasing the number of outer iterations $it_o$, where $it_o =
0$ represents the conventional receiver structure in which the
signal detector and the FEC decoder are serially concatenated, but
operate without exchanging soft information. However, the attainable
improvement gradually becomes smaller and the performance achieved after
three outer iterations becomes almost the same as that of four outer
iterations. This implies that the AB-Log-PDA based IDD scheme
essentially converges after three outer iterations. A similar
convergence profile is also observed for the optimal Exact-Log-MAP
based IDD, although its performance is always marginally better than
that of the corresponding AB-Log-PDA based IDD scheme. Notably, both IDD
schemes considered achieve $\textrm{BER} = 10^{-5}$ at about
$E_b/N_0 = 1$ dB after three iterations.

\textit{3) Impact of Nakagami-$m$ fading parameter $m$}
\begin{figure}[tbp]
\centering
\includegraphics[width=3.6in]{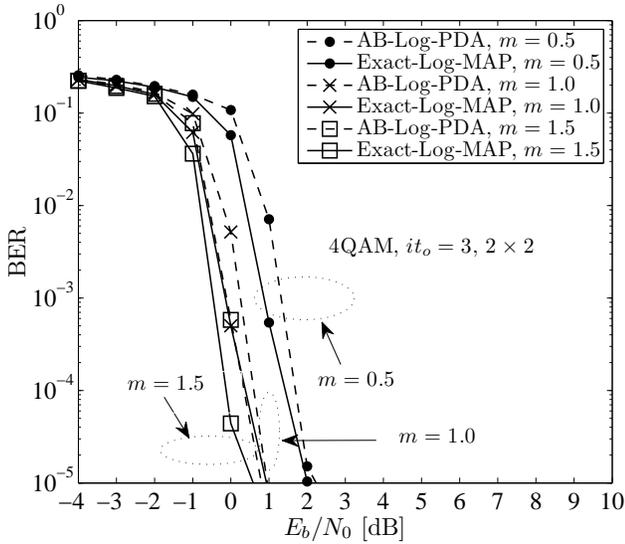}
\caption{Impact of Nakagami-$m$ fading parameter $m$ on the achievable BER of the AB-Log-PDA ($it_i = 0$) and the
Exact-Log-MAP based IDD schemes.} \label{fig:nakagami_m_on_BER}
\end{figure}
Fig. \ref{fig:nakagami_m_on_BER} shows the impact of different
$m$ values on the achievable BER performance
of the IDD schemes considered. As $m$ decreases, the achievable performance of both the IDD schemes
considered is degraded, since the fading
becomes more severe. However, the performance gap between the
AB-Log-PDA and the Exact-Log-MAP based IDD schemes is marginal for
all values of $m$ considered.

\textit{4) Impact of modulation order}
\begin{figure}[tbp]
\centering
\includegraphics[width=3.6in]{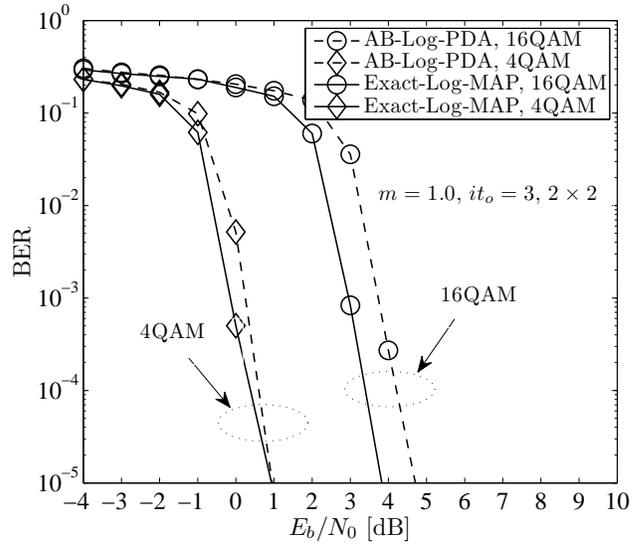}
\caption{Impact of the modulation order on the achievable BER of the
AB-Log-PDA ($it_i = 0$) and the Exact-Log-MAP based IDD schemes.}
\label{fig:modulation_order_on_BER}
\end{figure}
Additionally, in Fig. \ref{fig:modulation_order_on_BER} we
investigate the impact of the modulation order on the achievable
performance of the AB-Log-PDA based IDD scheme. It is observed that for higher-order
modulation, for example, 16QAM, the performance gap between the
AB-Log-PDA and the Exact-Log-MAP based IDDs becomes larger. This is
because the accuracy of the Gaussian approximation in the PDA
method degrades, when the modulation order is increased. More
specifically, as analyzed in Section
\ref{Sec:interference_plus_noise_distribution_analysis} and shown in
Fig. \ref{fig:GMM_PDF} as well as Fig.
\ref{fig:3D_GMM_after_iteration}, the interference-plus-noise term
${\bf{v}}_i$ obeys a multimodal Gaussian distribution associated with
$M^{N_t-1}$ Gaussian component-distributions. When $M$ is large,
there exist many interfering Gaussian component-distributions,
where the effect of each might be trivial, but their accumulated effect
may render the estimated $P(s_i = a_m|\bf{y})$ inaccurate, and hence inaccurate bit-based extrinsic LLRs might be
calculated using (\ref{eq:pseudo_symbol_APP_based_L_e}).

\textit{5) Impact of the number of transmit antennas}
\begin{figure}[tbp]
\centering
\includegraphics[width=3.6in]{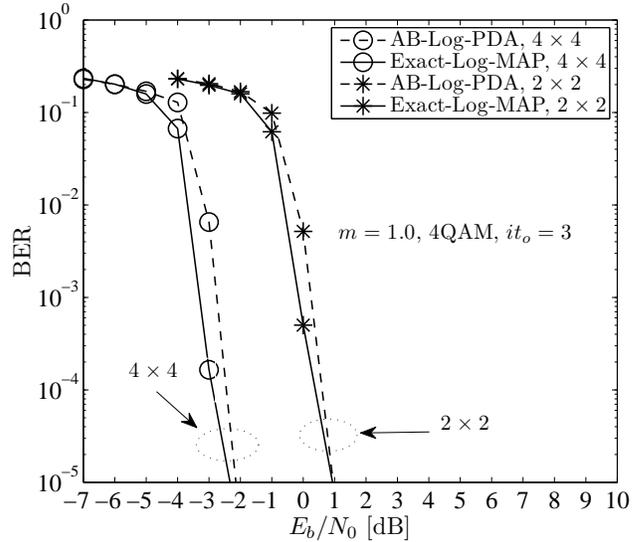}
\caption{Impact of the number of transmit antennas on the achievable
BER of the AB-Log-PDA ($it_i = 0$) and the Exact-Log-MAP based IDD schemes.}
\label{fig:transmit_antenna_on_BER}
\end{figure}
The impact of the number of transmit antennas $N_t$ on the
achievable performance of the AB-Log-PDA based IDD scheme is shown
in Fig. \ref{fig:transmit_antenna_on_BER}. On the one hand, upon increasing
$N_t$ (and $N_r$), an increased diversity gain is
obtained, and a more accurate Gaussian approximation is achieved
according to the central limit theorem. Hence we observe a
significant performance improvement, when moving from a $(2\times2)$-element
to a $(4\times4)$-element MIMO system. On the other hand, however, it
is observed that when
$N_t$ (and $N_r$) is increased, the performance gap
between the AB-Log-PDA and the Exact-Log-MAP based IDD receivers
is also increased. This is because the achievable diversity gain of the Exact-Log-MAP detector is higher
than that of the AB-Log-PDA detector, although a higher $N_t$
results in an improved Gaussian approximation quality.

\textit{6) Impact of channel-estimation error}
\begin{figure}[tbp]
\centering
\includegraphics[width=3.6in]{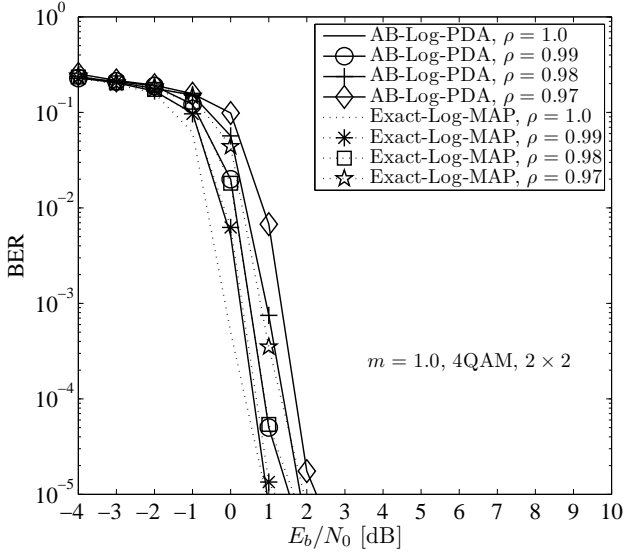}
\caption{Impact of channel-estimation error on the achievable
BER of the AB-Log-PDA ($it_i = 0$) and the Exact-Log-MAP based IDD schemes.}
\label{fig:CE_on_BER}
\end{figure}
Finally, the impact of the channel-estimation errors on the achievable BER performance of both the AB-Log-PDA and the Exact-Log-MAP based IDD schemes is investigated in Fig. \ref{fig:CE_on_BER}. The estimated channel matrix is given by $\hat{\bf{H}} = \rho{\bf{H}} +\sqrt{(1.0 - \rho^2)}\Delta{\bf{H}}$, where $0 \le \rho \le 1$ indicates the accuracy of channel-estimation. For example, $\rho = 1.0$ represents perfect channel-estimation and each entry of $\Delta{\bf{H}}$ obeys a zero-mean, unit-variance complex Gaussian distribution. It can be observed from Fig. \ref{fig:CE_on_BER} that the achievable BER performance of both IDD schemes is moderately degraded upon increasing the value of $\rho$ and that the AB-Log-PDA based IDD still achieves a performance similar to that of its  Exact-Log-MAP based IDD counterpart, even when the channel-estimation accuracy is as low as $\rho = 0.97$.

\subsection{Complexity Analysis}
Because the turbo codec module is common to both IDD schemes, and since we
have shown that both the AB-Log-PDA and the Exact-Log-MAP based IDD
schemes converge after three iterations in the scenarios considered,
the computational complexity of the proposed AB-Log-PDA based IDD
scheme can be evaluated by simply comparing its complexity to that
of the Exact-Log-MAP in a single iteration. As shown in Table
\ref{table:AB_Log_PDA}, the major computational cost of the
AB-Log-PDA per transmit symbol is the calculation of ${\bf{\Lambda
}}_i^{-1}$ and the matrix multiplication of
(\ref{eq:effective_metric}). Direct calculation of ${\bf{\Lambda
}}_i^{-1}$ imposes a computational cost of $\mathcal{O}(8N_r^3)$
real-valued operations (additions/multiplications), which is still
relatively expensive. Fortunately, by using the
Sherman-Morrison-Woodbury formula based ``speed-up'' techniques of
\cite{Luo:PDA_Sync_CDMA}, the computational cost of calculating
${\bf{\Lambda }}_i^{-1}$ can be reduced to
% $\mathcal{O}(4N_r^2)$
% real operations per transmit symbol, or
$\mathcal{O}(4N_tN_r^2)$
real-valued operations per iteration. Additionally, the calculation of
(\ref{eq:effective_metric}) requires
% $\mathcal{O}(4MN_r^2 + 2MN_r)$
% real operations per transmit symbol, or
$\mathcal{O}(4MN_tN_r^2 +
2MN_tN_r)$ real-valued operations per iteration. In summary, the
computational complexity of the AB-Log-PDA method is
$\mathcal{O}(4MN_tN_r^2 + 2MN_tN_r)$ + $\mathcal{O}(4N_tN_r^2)$ per
iteration.

\begin{figure}[t]
\centering
\includegraphics[width=3.6in, height = 3.1in]{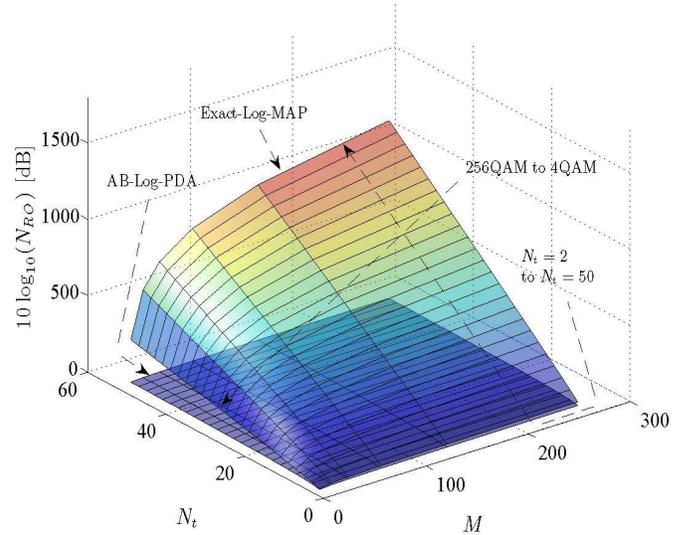}
\caption{Computational complexity comparison of the AB-Log-PDA ($it_i = 0$) and the Exact-Log-MAP algorithms in terms of the number of real
operations $N_{RO}$.}
\label{fig:complexity_comparison}
\end{figure}

By comparison, the Exact-Log-MAP algorithm has to calculate the
Euclidean distance ${\|{\bf{y}} - {\bf{Hs}}\|}^2$ for $M^{N_t}$ times,
hence its complexity order is $\mathcal{O}(M^{N_t})$. More
specifically, ${\|{\bf{y}} - {\bf{Hs}}\|}^2$ requires
$\mathcal{O}(4N_rN_t+6N_r)$ real-valued operations. Therefore, the
Exact-Log-MAP algorithm has a computational complexity of
$\mathcal{O}[M^{N_t}(4N_rN_t+6N_r)]$ real-valued operations per iteration,
which is significantly higher than that of the AB-Log-PDA,
especially when $N_t$, $N_r$ and $M$ have large values. This observation is further
confirmed by the results of Fig. \ref{fig:complexity_comparison}, where the computational complexity
of the two algorithms is compared in terms of the number of real-valued operations $N_{RO}$, while considering the scenario of
$N_r = N_t$ as an example. To elaborate a little further, the upper surface and the lower surface represent the computational complexity of the Exact-Log-MAP algorithm and of the proposed AB-Log-PDA algorithm, respectively. Since we assume $N_r = N_t$, the computational complexity of both algorithms becomes a function of the number of transmit antennas $N_t$ and of the modulation order $M$. We can observe from Fig. \ref{fig:complexity_comparison} that upon increasing $N_t$ and/or $M$, the computational complexity of the Exact-Log-MAP algorithm increases substantially faster than that of the AB-Log-PDA algorithm. Additionally, compared to $M$, $N_t$ plays a more significant role in determining the computational complexity of the two algorithms.

Finally, compared to the popular list sphere decoding (LSD) algorithm \cite{Hochwald:SD_near_capacity}, which has an exponential complexity lower bound,  especially for low SNR values \cite{Jalden:SD_complexity_journal}, the proposed AB-Log-PDA has the distinct advantage of a polynomial-time complexity (roughly a \textit{cubic} function of $N_t$, as shown above) for all SNR values. Although there exist other reduced-complexity variants of LSD, such as the list fixed-complexity sphere-decoder (LFSD) \cite{Barbero:soft_fixed_complexity_SD} and the soft $K$-best sphere-decoder using an improved ``look-ahead path metric'' \cite{Choi:sphere_decoder_look_ahead}, in general they still have a higher complexity than the AB-Log-PDA algorithm if the SNR value is low and/or the problem size (i.e. $N_t$ and $M$) is large. This is because finding the closest point in lattices is an NP-hard problem \cite{Agrell:closest_point_search_in_lattice}. To be more specific, the computational complexity of the $K$-best SD of \cite{Choi:sphere_decoder_look_ahead} is indeed reduced, but it remains of similar order to that of the LFSD. which is on the order of $\mathcal{O}(M^{\sqrt N_t})$ \cite{Jalden:FCSD_error_prob}.

\section{Conclusions}
\label{Sec:conclusions}
We demonstrate that the classic candidate-search based method of calculating bit-based extrinsic LLRs is not applicable to the family of PDA-based detectors. Additionally, in stark contrast to the existing literature, we demonstrate that the output symbol probabilities of the existing PDAs are not the true APPs, they are rather constituted by the normalized symbol likelihoods. Hence, surprisingly, the classic relationship, where the extrinsic LLRs are given by subtracting the \textit{a priori} LLRs from the \textit{a posteriori} LLRs does not hold for the existing PDA-based detectors, when the output probabilities of the existing PDAs are interpreted as APPs to generate \textit{a posteriori} LLRs. Based on these insights, we conceive the AB-Log-PDA method and identify the technique of calculating the bit-based extrinsic LLRs for the AB-Log-PDA, which results in a  simplified IDD receiver structure. Additionally, we demonstrate that we may completely dispense with any inner iterations within the AB-Log-PDA in the context of IDD receivers. Our complexity analysis and numerical results recorded for transmission over Nakagami-$m$ fading MIMO channels demonstrate that the proposed AB-Log-PDA based IDD scheme is capable of achieving a comparable performance to that of the optimal MAP detector based IDD receiver, while imposing a significantly lower computational complexity in the scenarios considered.

\ifCLASSOPTIONcaptionsoff
  \newpage
\fi

\bibliography{../../../../shaoshi_bib/IEEEabrv,../../../../shaoshi_bib/shaoshi_reference}
\newpage

\begin{IEEEbiography}[{\includegraphics[width=1in,height=1.25in,clip,keepaspectratio]{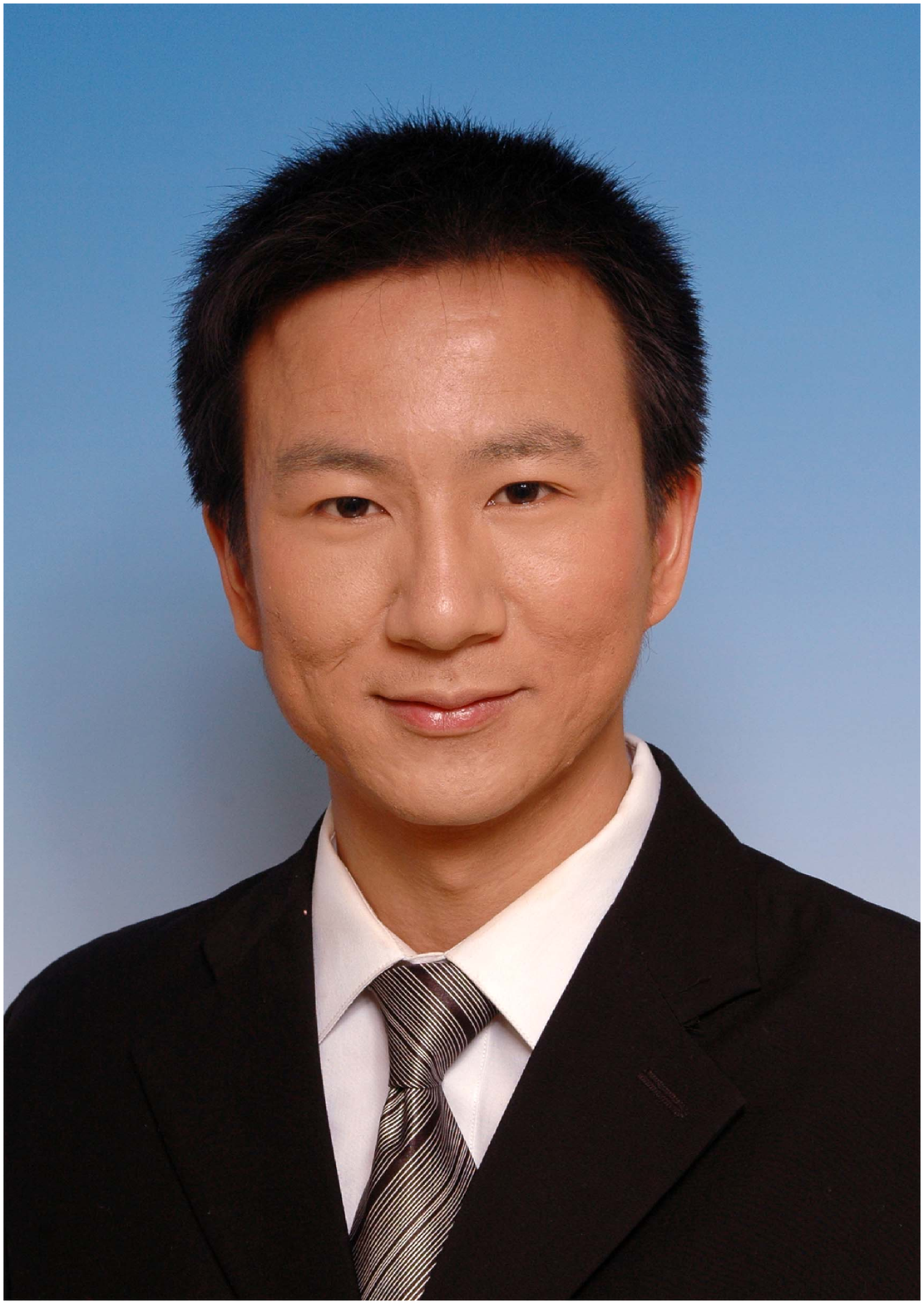}}] {Shaoshi Yang}
(S'09) received the B.Eng. Degree in information engineering from Beijing University of Posts and Telecommunications, Beijing, China, in 2006. He is currently working toward the Ph.D. degree in wireless communications with the School of Electronics and Computer Science, University of Southampton, Southampton, U.K., through scholarships from both the University of Southampton and the China Scholarship Council.

From November 2008 to February 2009, he was an Intern Research Fellow with the Communications Technology Laboratory, Intel Labs China, Beijing, where he focused on Channel Quality Indicator Channel design for mobile WiMAX (802.16 m). His research interests include multiuser detection/multiple-input mutliple-output detection, multicell joint/distributed processing, cooperative communications, green radio, and interference management. He has published in excess of 20 research papers on IEEE journals and conferences. 

Shaoshi is a recipient of the PMC-Sierra Telecommunications Technology Scholarship, and a Junior Member of the Isaac Newton Institute for Mathematical Sciences, Cambridge, UK. He is also a TPC member of both the 23rd Annual IEEE International Symposium on Personal, Indoor and Mobile Radio Communications (IEEE PIMRC 2012), and of the 48th Annual IEEE International Conference on Communications (IEEE ICC 2013). 
\end{IEEEbiography}

\begin{IEEEbiography}[{\includegraphics[width=1in,height=1.25in,clip,keepaspectratio]{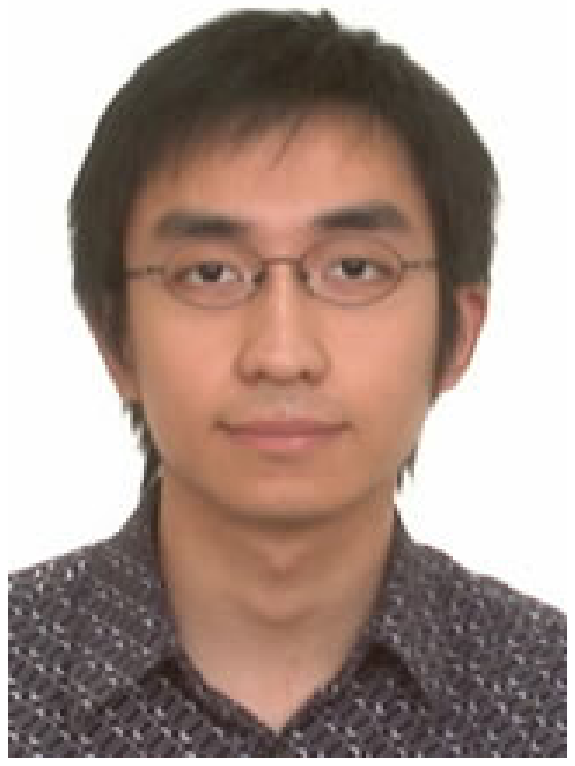}}] {Li Wang}
(S'09-M'10) was born in Chengdu, China, in 1982. He received his BEng degree in Information Engineering from Chengdu University of Technology (CDUT), Chengdu, China, in 2005 and his MSc degree with distinction in Radio Frequency Communication Systems from the University of Southampton, UK, in 2006. Between October 2006 and January 2010 he was pursuing his PhD degree in the Communications Group, School of Electronics and Computer Science, University of Southampton, and meanwhile he participated in the Delivery Efficiency Core Research Programme of the Virtual Centre of Excellence in Mobile and Personal Communications (Mobile VCE). 

Upon completion of his PhD in January 2010 he conducted research as a Research Fellow in the School of Electronics and Computer Science at the University of Southampton. During this period he was involved in Project \#7 of the Indian-UK Advanced Technology Centre (IU-ATC): advanced air interface technique for MIMO-OFDM and cooperative communications. In August 2012 he joined the R\&D center of Huawei Technologies in Stockholm, Sweden, working as Senior Engineer of Baseband Algorithm Architecture. He has published over 30 research papers in IEEE/IET journals and conferences, and he also co-authored one John Wiley/IEEE Press book. He has broad research interests in the field of wireless communications, including PHY layer modeling, link adaptation, cross-layer system design, multi-carrier transmission, MIMO techniques, CoMP, channel coding, multi-user detection, non-coherent transmission techniques, advanced iterative receiver design and adaptive filter.
\end{IEEEbiography}

\begin{IEEEbiography}[{\includegraphics[width=1in,height=1.25in,clip,keepaspectratio]{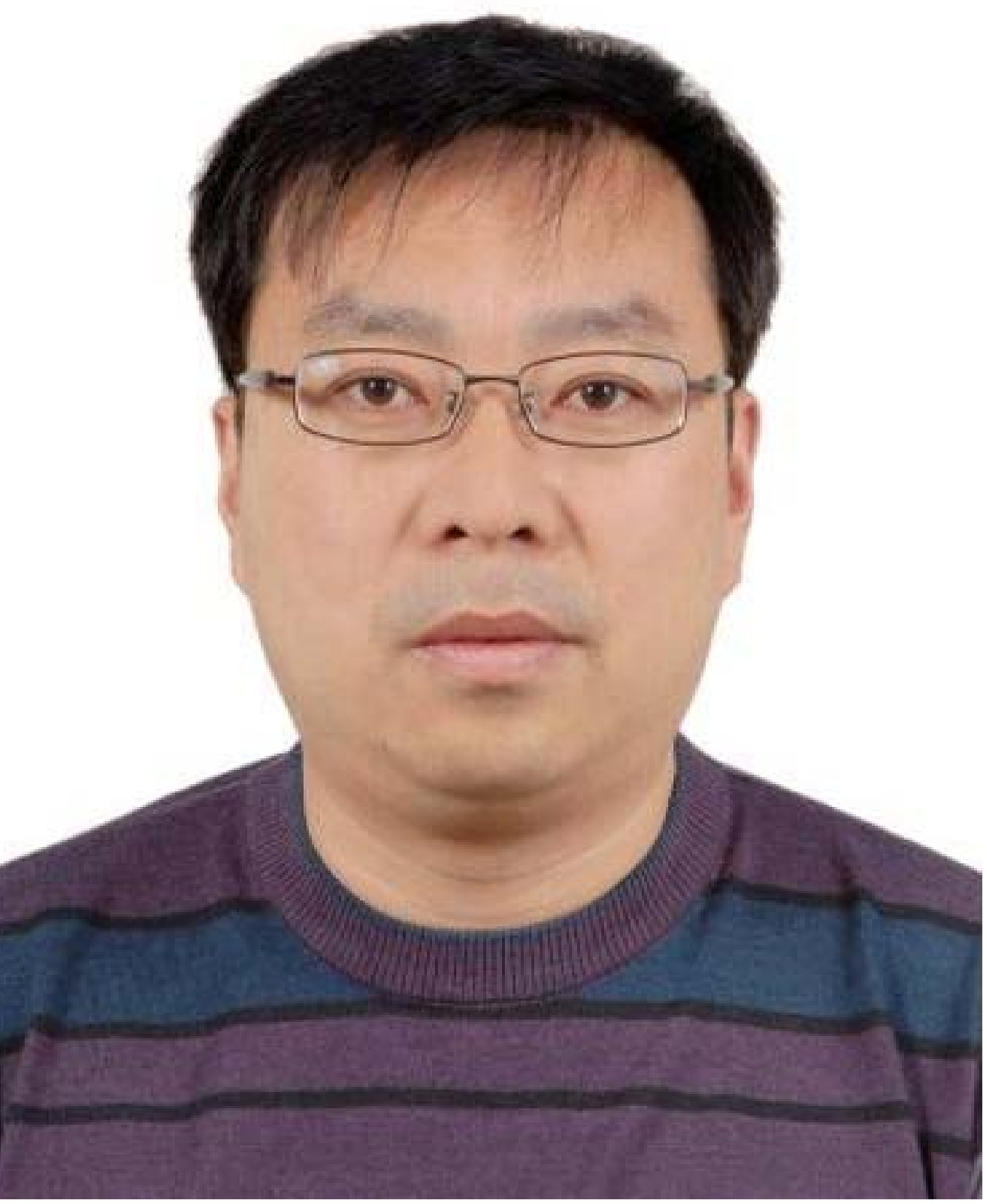}}] {Tiejun Lv}
(M'08-SM'12) received the M.S. and
Ph.D. degrees in electronic engineering from the University of Electronic Science and Technology
of China, Chengdu, China, in 1997 and 2000, respectively. From January 2001 to December 2002,
he was a Postdoctoral Fellow with Tsinghua University, Beijing, China. From September 2008 to
March 2009, he was a Visiting Professor with the Department of Electrical Engineering, Stanford University, Stanford, CA. He is currently a Professor with the School of Information and Communication Engineering, Beijing University of Posts and Telecommunications. He is the author of more than 100 published technical papers on the physical layer of
wireless mobile communications. His current research interests include signal
processing, communications theory and networking. Dr. Lv is also a Senior
Member of the Chinese Electronics Association. He was the recipient of the
"Program for New Century Excellent Talents in University" Award from the
Ministry of Education, China, in 2006.
\end{IEEEbiography}

\begin{IEEEbiography}[{\includegraphics[width=1in,height=1.25in,clip,keepaspectratio]{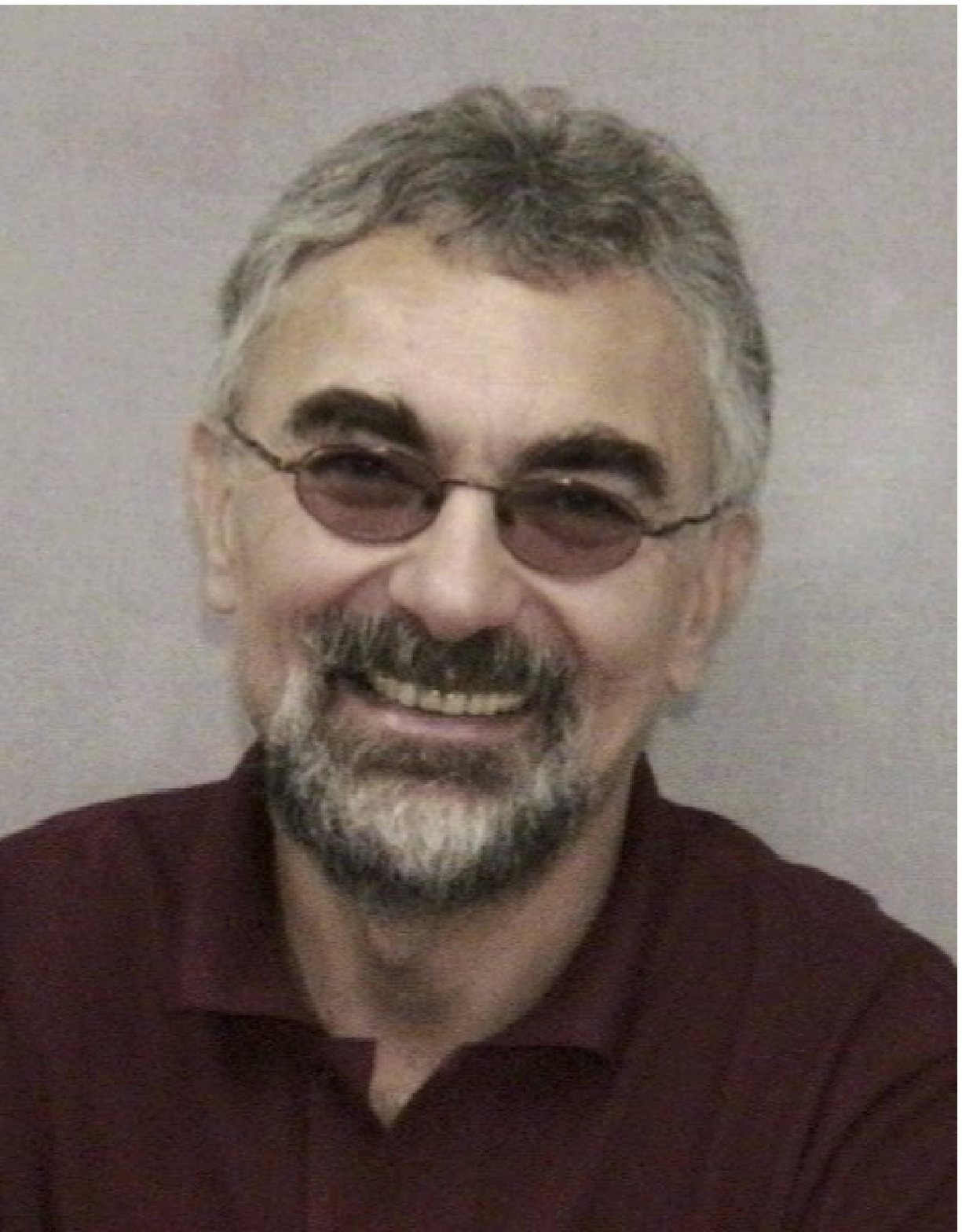}}] {Lajos Hanzo}
(F'04) received his degree in electronics in
1976 and his doctorate in 1983. In 2004 he was awarded the Doctor of Sciences (DSc) degree by University of Southampton, U.K., and in 2009 he was awarded the honorary doctorate ``Doctor Honoris Causa'' by the Technical University of
Budapest. 

During his 36-year career in telecommunications he has held
various research and academic posts in Hungary, Germany and the
UK. Since 1986 he has been with the School of Electronics and Computer
Science, University of Southampton, U.K., where he holds the chair in
telecommunications. He has successfully supervised 80 PhD students,
co-authored 20 John Wiley/IEEE Press books on mobile radio
communications totalling in excess of 10000 pages, published 1300
research entries at IEEE Xplore, acted both as TPC and General Chair
of IEEE conferences, presented keynote lectures and has been awarded a
number of distinctions. Currently he is directing a 100-strong
academic research team, working on a range of research projects in the
field of wireless multimedia communications sponsored by industry, the
Engineering and Physical Sciences Research Council (EPSRC), U.K., the
European IST Programme and the Mobile Virtual Centre of Excellence
(VCE), U.K.. He is an enthusiastic supporter of industrial and academic
liaison and he offers a range of industrial courses. 

Dr. Hanzo is a Fellow of the Royal Academy of Engineering, and the Institution of Engineering and Technology (IET), as well as the European Association for Signal Processing (EURASIP). He is also a Governor of the IEEE VTS. During 2008-2012 he was the
Editor-in-Chief of the IEEE Press and a Chaired Professor at
Tsinghua University, Beijing. His research is also funded by the
European Research Council's Senior Research Fellow Grant. For further
information on research in progress and associated publications please
refer to http://www-mobile.ecs.soton.ac.uk
\end{IEEEbiography}
\end{document}